\documentclass[12pt]{article}
\newcommand{\be}{\begin{equation}}
\newcommand{\ee}{\end{equation}}
\newcommand{\bea}{\begin{eqnarray}}
\newcommand{\eea}{\end{eqnarray}}
\newcommand{\ba}{\begin{array}}
\newcommand{\ea}{\end{array}}

\def\bbox{{\,\lower0.9pt\vbox{\hrule \hbox{\vrule height 0.2 cm
\hskip 0.2 cm \vrule height 0.2 cm}\hrule}\,}}
\newcommand{\dsl}{\pa \kern-0.5em /}

\newcommand{\nn}{\nonumber \\}

\font\mybb=msbm10 at 10pt
\def\bb#1{\hbox{\mybb#1}}

\def\bR {\bb{R}}

\usepackage{epsfig}

\begin{document}



\begin{titlepage}
\rightline{DAMTP-2005-94}
\rightline{\tt{hep-th/0510115}}

\bigskip

\begin{center}
\baselineskip=16pt
{\Large\bf  Dilaton Domain Walls and}
{\Large\bf 
 Dynamical Systems}
\vskip 0.3cm
{\large {\sl }}
\vskip 10.mm
{\bf ~Julian Sonner  and Paul K. Townsend}
\vskip 1cm
{\small
Department of Applied Mathematics and Theoretical Physics,\\
University of Cambridge, \\
Centre for Mathematical Sciences, \\
Wilberforce Road, \\
Cambridge CB3 0WA, UK
}

\end{center}
\bigskip

\par
\begin{center}
{\bf ABSTRACT}
\end{center}
\begin{quote}

Domain wall solutions of $d$-dimensional gravity coupled to a dilaton 
field $\sigma$ with an exponential potential $\Lambda e^{-\lambda\sigma}$ 
are shown to be governed by an autonomous  dynamical system,  
with a transcritical bifurcation as a function of the parameter 
$\lambda$ when $\Lambda<0$. All phase-plane trajectories are found 
exactly for $\lambda=0$, including separatrices corresponding to 
walls that interpolate between $adS_d$ and $adS_{d-1} \times\bR$, 
and the exact solution is found for $d=3$. 
Janus-type solutions are interpreted as marginal bound states
of these ``separatrix walls''. All flat domain wall solutions, which are given 
exactly for any $\lambda$, are shown to be supersymmetric for some 
superpotential $W$, determined by the solution.


\end{quote}

\end{titlepage}

\section{Introduction}
\setcounter{equation}{0}

There are many supergravity models of interest for which the action 
can be consistently truncated to a $d$-dimensional action for metric 
$g_{\mu\nu}$ and dilaton field $\sigma$ with Lagrangian density
\be\label{origlag}
{\cal L} = \sqrt{-\det g}\left[ R - {1\over2} (\partial\sigma)^2 
- \Lambda e^{-\lambda\sigma}\right]\, , 
\ee
where $\lambda$ is a constant, which we may assume to be non-negative, 
and $\Lambda$ is a non-zero constant, of either sign\footnote{A shift 
of $\sigma$ has the effect of scaling $\Lambda$ when $\lambda\ne0$, 
so  only the sign of $\Lambda$ is physically relevant,  
but the magnitude of  $\Lambda$ is physically relevant when 
$\lambda=0$.}, that equals the cosmological constant when $\lambda=0$.  
Cosmological solutions for this class of  models  have been much 
studied, and are well-understood;  in particular, a qualitative 
understanding of the entire space of solutions for a given $\lambda$ 
is made possible by the observation that the equations governing 
homogeneous and isotropic cosmologies define a 2-dimensional 
autonomous dynamical system \cite{Halliwell:1986ja}. 

Domain wall solutions of the same class of models have also 
attracted considerable attention, in part because there exist 
domain wall solutions that, in a supergravity context, preserve 
some fraction of supersymmetry\footnote{The issue of supersymmetry
preservation for dilaton domain walls appears to have been first adressed 
in \cite{Cvetic:1993yq}, albeit in a more general model with domain
wall solutions that differ from those discussed here.}. 
The equations to be solved for 
domain walls are formally rather similar to those for cosmologies;  
instead of an evolution in time one has an `evolution' in a space 
coordinate, and one `evolves' a $(d-1)$-dimensional spacetime 
instead of a spatial hypersurface. In this paper we exploit this 
similarity to show that the equations for domain wall solutions 
can also be expressed as those of a 2-dimensional autonomous dynamical 
system. This again allows a  qualitative understanding of 
the {\it entire} space of domain wall solutions, for  given 
$\lambda$, as a set of phase-plane 
trajectories.  

The trajectories corresponding to flat domain walls (with a Minkowski 
worldvolume geometry) divide the phase plane into three regions. For 
$\Lambda<0$, two regions for which the wall's worldvolume geometry is 
de Sitter ($dS$) are separated by one for which  it is anti-de Sitter
($adS$), and {\it vice versa} for  $\Lambda>0$. All flat domain wall 
solutions can be found exactly \cite{Lu:1995hm, Lu:1996hh}; here we 
recover these results following the method used in
\cite{Townsend:2003qv} to find all flat cosmological solutions 
(first given in \cite{Burd:1988ss}). The qualitative behaviour of all 
other trajectories, corresponding to walls with $dS$ or $adS$ worldvolume 
geometry, is determined by the positions and the nature of the fixed 
points. The analysis  is essentially the same as  the cosmological 
case but the  spacetime interpretation of the solutions is of course 
different.  

The $\lambda=0$ case is special, and of particular interest in that 
the full spacetime, and not just the domain wall's worldvolume, 
can be de Sitter or anti-de Sitter. For this case, we find {\it all}
phase-plane trajectories exactly. These include  two fixed points that
each correspond to the  $adS_d$ vacuum, foliated by Minkowski spaces.
Other trajectories correspond to the same $adS_d$ vacuum but foliated by
$dS$ or $adS$ spaces. The trajectory corresponding to the $adS_{d-1}$ 
foliation of $adS_d$  is actually a special case of a one-parameter 
family of trajectories that interpolate between the two $adS_d$ fixed 
points. These correspond to the ``Janus'' solutions of 
\cite{Freedman:2003ax}. A limit of these Janus trajectories yields 
the union of two separatrix trajectories, each interpolating between 
one of the $adS_d$ fixed points and one of two other fixed points, 
each of which corresponds to an $adS_{d-1}\times \bR$ solution (the
``curious linear dilaton'' solution of \cite{Freedman:2003ax}). 
These separatrices correspond to new solutions, which we call 
``separatrix walls'', and we find the exact separatrix wall solution
for $d=3$. We point out that Janus solutions can be interpreted as 
marginal bound states of separatrix walls. 

For $\lambda>0$ our results are more qualitative, although all
fixed-point (in addition to flat) domain wall solutions can be found 
exactly, and all trajectories can be found exactly in the
$\lambda\to\infty$ limit. One interesting feature of the family 
of $\Lambda<0$ phase plane trajectories, parametrized by $\lambda$, is that 
a bifurcation occurs at a critical value $\lambda_c$ of $\lambda$. 
Here we show that this is, in the language of dynamical systems, 
a {\it transcritical}  bifurcation. Another interesting feature is
that for $\lambda<\lambda_c$ there is a one-parameter family of 
Janus-type solutions that are similarly  ``two-faced''  but which are 
asymptotic to 
a $\lambda$-deformation of $adS_d$.  

As already mentioned, one reason for interest in domain wall solutions
of our model is that, in the supergravity context,  
domain walls may preserve some fraction of the supersymmetry of 
the supergravity vacuum. This issue has been investigated 
previously in the context of various specific supergravity models 
that have a consistent truncation to (\ref{origlag}); we will
comment later on how this work fits in with our results, 
which are model independent  in the following sense. On general  
grounds  \cite{Boucher:1984yx,Townsend:1984iu,Skenderis:1999mm} 
one expects the potential $V(\sigma)$ for 
any single scalar field to take the form 
\be\label{VWeq}
V= 2 \left[(W')^2 -\alpha^2 W^2\right]  \qquad \qquad 
\left(W' \equiv dW/d\sigma\right)\, ,
\ee
for some super-potential function $W(\sigma)$. Locally, 
one can view this as a differential 
equation that determines $W$ for given $V$ \cite{DeWolfe:1999cp}, 
but the global situation is more subtle. 
In our case, for which $V= \Lambda e^{-\lambda\sigma}$, we will see that
(\ref{VWeq}) does not always determine a unique superpotential, so 
that it is possible for a domain wall solution to be supersymmetric 
for one choice of superpotential and not for another.
We shall say that a domain wall solution is ``supersymmetric''  
if it preserves supersymmetry for {\it some} choice of 
superpotential. Remarkably, we find that the possible 
superpotentials correspond to the possible {\it flat} domain 
wall solutions, and that {\it any} flat domain wall is 
supersymmetric for a choice of superpotential that is actually 
determined by the solution! 

Essentially the same point is made 
in \cite{Freedman:2003ax}, as we learnt after submission to 
the archives of an earlier version of this paper. The conclusions 
of \cite{Freedman:2003ax} on the supersymmetry of flat walls 
apply for a general scalar potential $V$, but those presented here 
for exponential potentials are more complete. For curved walls, 
we find that the only ``supersymmetric'' solutions are the $dS_{d-1}$ 
and $adS_{d-1}$ foliations of $adS_d$, but a larger class of 
curved ``supersymmetric'' domain walls is found in
\cite{Freedman:2003ax} by allowing for a  matrix-valued 
superpotential. We comment further on this later but 
otherwise leave to the reader any more detailed comparison with 
\cite{Freedman:2003ax}. 

We begin by introducing the constants
\be
\alpha = \sqrt{(d-1)\over 2(d-2)}\, , \qquad \beta = 
{1\over \sqrt{2(d-1)(d-2)}}\, . 
\ee
Now consider the domain wall ansatz
\be
ds^2_d = e^{2\alpha \varphi} f^2(z) dz^2 + 
e^{2\beta\varphi} d\Sigma_k^2\, ,\qquad \sigma= \sigma(z)
\ee
for arbitrary function $f(z)$, where $d\Sigma_k^2$ is the metric
of a $(d-1)$-dimensional homogeneous spacetime with inverse 
radius of curvature equal to $k$; the scalar curvature is therefore 
$k(d-1)(d-2)$. As in the cosmological case, we may restrict to
$k=-1,0,1$ without loss of generality. The isometry group is $SO(d,1)$ 
for $k=1$, $ISO(d-1)$ for $k=0$, and $SO(d-2,2)$ for $k=-1$. 
Thus, $d\Sigma_k^2$ is the metric of a ``unit-radius'' de Sitter space 
for $k=1$, a Minkowski space for $k=0$ and a ``unit-radius'' anti-de Sitter 
space for $k<0$; note that the domain wall is flat for $k=0$ but 
curved for $k\ne0$. This ansatz yields the effective 
Lagrangian\footnote{Apart from a minor change of notation, and the 
interpretation of the independent variable, this is identical to 
the Lagrangian obtained in \cite{Bergshoeff:2005bt}  for the 
cosmological case; its domain-wall interpretation appeared in 
unpublished notes of E. Bergshoeff, A. Collinucci and D. Roest 
that were preliminary to that work. For either interpretation, 
it can be verified directly that solutions of the effective 
Lagrangian yield solutions of the equations of motion of (\ref{origlag}).}
\be
L= {1\over2 f}\left[ \dot\varphi^2 - \dot\sigma^2\right] + 
f\left[k(d-1)(d-2) e^{\varphi/\alpha} - 
\Lambda e^{2\alpha\varphi - \lambda \sigma} \right]\, ,
\ee
where the overdot indicates a derivative with respect to $z$. This 
can be interpreted as a reparametrization-invariant Lagrangian for 
a relativistic particle with a `time'-dependent potential energy in 
a 2-dimensional Minkowski spacetime. 

\section{The dynamical system}
\setcounter{equation}{0}

If we fix the $z$-reparametrization invariance by choosing
\be\label{fchoice}
f(z) = e^{\lambda\sigma/2 - \alpha\varphi}\, , 
\ee
then the equations of motion for $(\varphi,\sigma)$ and $f$ 
become equivalent to the equations
\bea\label{ds}
\ddot\sigma &=& {1\over2}\lambda \dot\sigma^2 - 
\alpha \dot\varphi \dot\sigma - \lambda \Lambda\nn
\ddot\varphi &=& {1\over2}\lambda \dot\sigma\dot\varphi - 
\beta \dot\varphi^2 - {1\over 2\alpha} \dot\sigma^2 -  2\beta \Lambda \, ,
\eea 
together with the constraint 
\be\label{con}
\dot\varphi^2 - \dot\sigma^2 + 2\Lambda =  
{k\over \beta^2} e^{\lambda\sigma -2\beta\varphi} \, .
\ee
We note here that the domain wall metric for the choice (\ref{fchoice}) is
\be\label{metricchoice}
ds^2_d = e^{\lambda\sigma(z)} dz^2 + e^{2\beta \varphi(z)} d\Sigma_k^2\, . 
\ee

Equations (\ref{ds}) define a 2-dimensional  autonomous dynamical 
system, with coordinates $(\dot\sigma,\dot\varphi)$.  The entire 
space of phase-plane trajectories is determined by the
positions and nature of the fixed points. For $k=0$ the constraint 
(\ref{con}) becomes the 
 hyperbola
\be
\dot\varphi^2 - \dot\sigma^2 =  -2\Lambda\, , 
\ee
and the two branches of this hyperbola divide the phase plane into 
three regions. For $\Lambda<0$ there is a `central'  $k=-1$ region 
containing the line $\dot\varphi=0$ that separates two $k=1$ regions. 
For $\Lambda>0$ there is a central $k=1$ region containing the line 
$\dot\sigma=0$ that separates two $k=-1$ regions. For $k\ne 0$ the 
constraint merely determines the value of $\lambda\sigma-
2\beta\varphi$ at a given point on a phase-plane trajectory, so it 
has no effect on the trajectories themselves. We shall therefore 
concentrate on the equations (\ref{ds}). In the notation
\be
u= \dot\sigma\, \qquad v= \dot\varphi\, , 
\ee
these two equations become
\bea\label{DS}
\dot u &=& {1\over2}\lambda u^2 - \alpha uv - 
\lambda \Lambda\nonumber\\
\dot v &=& {1\over2}\lambda uv - \beta v^2 - 
{1\over 2\alpha} u^2 -2\beta\Lambda\, .
\eea
The autonomous dynamical system defined by these equations differs 
from the one that governs FLRW cosmologies only by a flip of the 
signs of $\Lambda$ and $k$. Thus, the domain wall trajectories for 
negative $\Lambda$ are the same as the cosmological trajectories 
for positive $\Lambda$, and vice-versa, but with the opposite sign 
of $k$ in each case. 

\subsection{Fixed points}

As with any autonomous dynamical system, the first task is to identify 
the fixed points, in this case the points in the $(u,v)$ plane at
which $(\dot u, \dot v)=(0,0)$. Following the analysis of  
\cite{Townsend:2004zp} for the cosmological case, we proceed from 
the observation that the fixed points are such that
\be
\left(\lambda u -2\beta v\right)\left(\lambda v- 2\alpha u\right)=0\, . 
\ee
There are thus two types of fixed point:
\begin{itemize}

\item Type 1: $u= {\lambda \over 2\alpha}v$. In this case the 
fixed point conditions can be satisfied only if $\lambda \ne
2\alpha$. 
Given this, one finds that 
\be
v^2 = {8\alpha^2 \over \lambda^2 - 4\alpha^2} \, \Lambda\, , 
\ee
which implies that a fixed point exists for $\Lambda <0$ iff 
$\lambda<2\alpha$ and for $\Lambda>0$ iff $\lambda >2\alpha$. 
Thus, at this type of fixed point, 
\be
u= \pm \lambda K(\lambda)\, , \qquad v= \pm 2\alpha K(\lambda) \, , 
\ee
where
\be\label{klambda}
K(\lambda)=  \sqrt{\left|{2\Lambda \over \lambda^2 -
    4\alpha^2}\right|}\, . 
\ee
It follows that
\be
v^2-u^2 = 2|\Lambda|
\ee 
at the fixed point, and hence that that these fixed points lie on the 
$k=0$ hyperbola; in fact, these fixed points come in pairs, one on 
each branch of the  $k=0$ hyperbola.  

\item Type 2:  $v={\lambda \over 2\beta}u$. In this case the fixed 
point conditions are satisfied when
\be
u^2 = -{2\over d-2}\Lambda\, ,
\ee
which shows that $\Lambda$ must be negative. Thus,  at this type 
of fixed point, 
\be
u= \pm \lambda_c \sqrt{|\Lambda|} ,\qquad 
v= \pm \lambda \sqrt{(d-1)|\Lambda|}
\ee
where
\be
\lambda_c = \sqrt{2\over d-2}\, . 
\ee
Again, each fixed point of this type occurs in pairs, one on each of 
the lines $|u|=\lambda_c\sqrt{|\Lambda|}$. The constraint  (\ref{con}) 
now yields
\be
\left(\lambda^2 - \lambda_c^2\right)|\Lambda| 
=2(d-2)ke^{\lambda\sigma-2\beta\varphi}\, , 
\ee
which shows that $k=-1$ for $\lambda<\lambda_c$ and $k=1$ 
for $\lambda>\lambda_c$. For $\lambda=\lambda_c$ this fixed 
point coincides with the one on the $k=0$ hyperbola.

\end{itemize}

To summarize, fixed points occur for $\Lambda>0$ only if 
$\lambda>2\alpha$, and then only
for $k=0$, with one fixed point on each branch of the $k=0$ 
hyperbola. A similar pair of $k=0$ fixed points occurs for 
$\Lambda<0$ when $\lambda<2\alpha$, but there is also another 
pair of fixed points, with $k=-1$ for $\lambda<\lambda_c$ and 
$k=1$ for $\lambda>\lambda_c$. For $\lambda=\lambda_c$, each 
of these  fixed points coincides with one of the pair of $k=0$ 
fixed points. This implies a bifurcation 
in the family of dynamical systems parametrized by $\lambda$. 
We now turn to an investigation of the nature of this bifurcation. 

\subsection{The transcritical bifurcation}

We now concentrate on the case of $\Lambda<0$, and for simplicity 
we set  $\Lambda=-1$.  
As we are interested in what happens when $\lambda \approx 
\lambda_c$, we define 
\be
\mu = \lambda-\lambda_c
\ee
as a new parameter. Also, for later convenience, we define 
\be
J(\mu) \equiv  \sqrt{(d-2)(d-10)\lambda^2 + 16} = (d-2)\lambda_c + 
{\cal O}\left(\mu\right)\, . 
\ee

To investigate the nature of the bifurcation at $\mu=0$, it is 
convenient to introduce the 
shifted variables 
\be
u= \lambda_c + \tilde u\, ,\qquad
v= \sqrt{d-1}\, \lambda + \tilde v\, . 
\ee
The $k=-1$ fixed point in the $u>0$ half-plane is now at 
$(\tilde u,\tilde v)=(0,0)$ for any value of 
$\mu$. To put the equations into a standard form in a neigbourhood 
of this fixed point, we 
introduce the new variables
\be 
x= \tilde v +s_- \tilde u\, ,\qquad 
y= \tilde v + s_+ \tilde u \, ,
\ee
where
\be
s_\pm = {1\over 4\alpha}\left[ (d-4)\lambda \pm J \right]\, . 
\ee

\vskip2em
\begin{figure}[!h]
  \begin{center}
    \epsfig{file=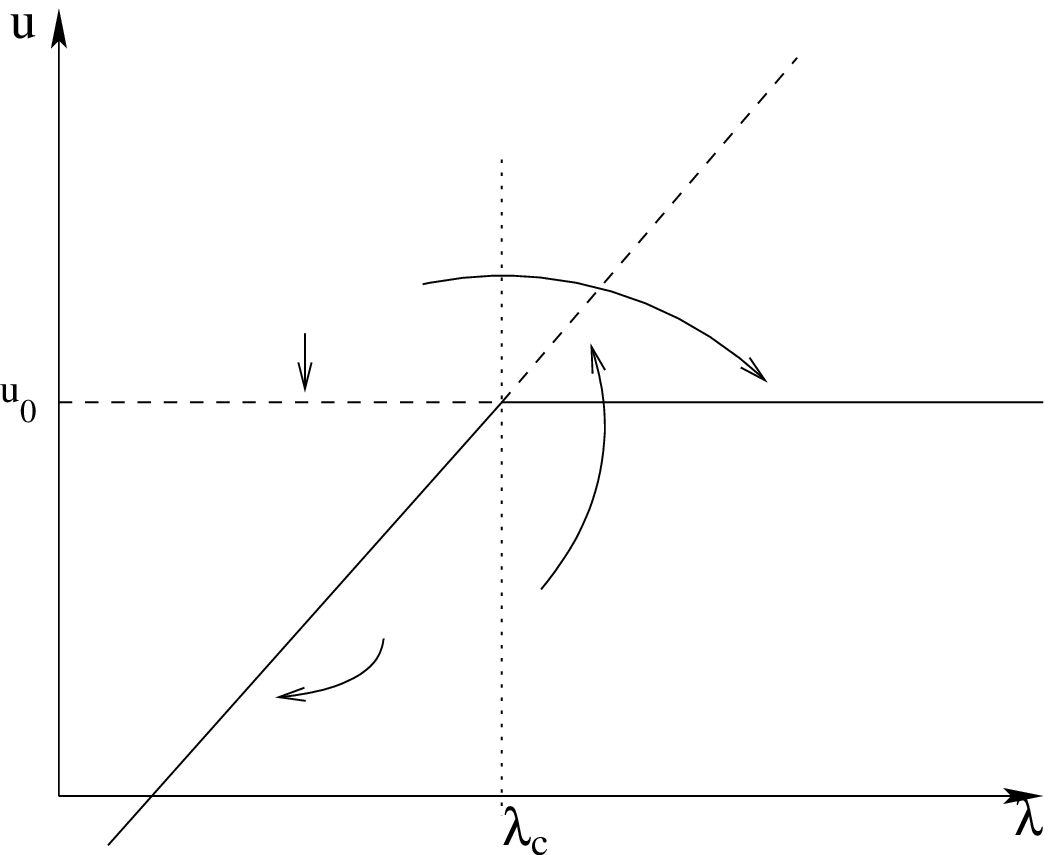,width=6.5cm}\hskip3em 
\begin{picture}(2,2)(0,0)
      \put(-150,150){ $\lambda \sim \lambda_c$,}
      \put(-210,100){\scriptsize{$k\ne 0$ fixed point}}
      \put(-150,40){\scriptsize{$k= 0$ fixed point}}
      \end{picture}
     \end{center}\caption{\small{Bifurcation Diagram. 
       The transcritical bifurcation
      corresponds to an exchange of stable (solid line) and unstable
      (dashed line) directions between two fixed points.}}
\label{Fig:bifurcation}
\end{figure}

One then finds that the equations (\ref{DS})  take the 
form\footnote{This is a 
standard form,  as given in (for example) Chapter 8 of
\cite{Glendinning}, which we found to be a useful reference.} 
\bea\label{evolution}
\dot x &=& A(\mu) \, x + F_1(x,y,\mu)\, , \nonumber\\
\dot y &=& -B(\mu)\,  y + F_2(x,y,\mu)\, ,
\eea
where $B$ is positive, $A(0)$ vanishes, and $(F_1,F_2)$ are 
two functions that both vanish and have vanishing first-derivatives 
with respect to $x,y$ and $\mu$ at $(x,y,\mu) =(0,0,0)$. 
Specifically, one finds that
\be
A= {1\over4}\lambda_c \left[ J-(d-2)\lambda_c\right]\, ,\qquad
B= {1\over4}\lambda_c \left[ J+(d-2)\lambda_c\right]\, , 
\ee
and that
\bea
F_1 &=& {\beta\over 2J^2}\Bigg\{ -\left[4(3d-1) -
(d+8)(d-2)\lambda^2 + 3(d-2)\lambda J\right] x^2 \nonumber\\
&& + \, \left[8(d-3) -(d-1)(d-6)(d-2)\lambda^2 + 
(d-1)(d-2)\lambda J\right] xy \nonumber\\
&& + \, \left[ 4(d-3) + (d-2)(d^2-10 d + 18)\lambda^2 -
(d-4)(d-2)\lambda J\right] y^2\Bigg\}\, ,
\nonumber\\
F_2 &=& {\beta\over 2J^2}\Bigg\{\left[ 4(d-3) + 
(d-2)(d^2-10d+18)\lambda^2 +(d-4)(d-2)\lambda J\right] x^2 \nonumber\\
&& +\, \left[8(d-3)-(d-1)(d-6)(d-2)\lambda^2 -
(d-1)(d-2)\lambda J\right] xy \nonumber\\
&& -\, \left[4(3d-1)-(d+8)(d-2)\lambda^2 -3(d-2)\lambda J\right] y^2\bigg\}\, . 
\eea
Expansion in powers of $\mu$ yields
\be
A(\mu) = -2\mu\lambda_c + {\cal O}\left(\mu^2\right) \, , \qquad
B(\mu) =  1 + {1\over2}(d-6)\mu\lambda_c + {\cal O}\left(\mu^2\right)\, , 
\ee
which confirms that the fixed point at the origin is hyperbolic for 
$\mu\ne0$ but non-hyperbolic for 
$\mu=0$, and 
\bea
F_1 &=&  \beta\left[-4x^2 + 4xy - y^2 \right] + 
{\cal O}\left(\mu\right)\ \nonumber \\
F_2 &=&  \beta\left[(d-5)x^2 -(d-5)xy - y^2 \right] +
{\cal O}\left(\mu\right)\, , 
\eea
where the ${\cal O}(\mu)$ terms in $(F_1,F_2)$  are  
quadratic in $(x,y)$, so that
the functions $(F_1,F_2)$ are $\mu$-independent to quadratic order. 

The behaviour near $\mu=0$ is determined by the dynamics on a 
2-dimensional  `extended  centre manifold', this being the 
centre manifold of the extended system in which $\mu$ is taken 
as a third variable with the trivial equation $\dot\mu=0$. The 
extended centre manifold is given by $y=h(x,\mu)$ for some 
function $h$ that both vanishes and has vanishing first-derivatives 
with respect to $x$ and $\mu$ at $(x,\mu)=(0,0)$. The function $h$ 
can be found as a power series in $x$ and $\mu$ by requiring 
consistency with the evolution equations (\ref{evolution}). This yields
\be
h(x,\mu) = (d-5)\alpha x^2 + \dots\, ,
\ee
where the dots indicate terms at least cubic in the two variables 
$(x,\mu)$; note that $h$ is independent of $\mu$ at quadratic order, 
which is all that we will need. Substitution of $y=h(x,\mu)$ into 
the equation for $x$ yields an equation of the form
\be\label{centredyn}
\dot x = G(x,\mu) \, , \qquad G= -2\mu\lambda_c x -4\beta x^2 + 
\dots \, .
\ee
In terms of the rescaled variable $w$ and the rescaled parameter 
$\nu$, defined by
\be
w= 4\beta x\, ,\qquad \nu = -2\mu\lambda_c\, , 
\ee
this equation takes the form
\be
\dot w = \nu w - w^2 + \dots \, . 
\ee
This is the standard form for a {\it transcritical}  bifurcation in 
which the stability properties of the fixed points are exchanged as 
they cross at $\nu=0$, as illustrated in the bifurcation diagram of
Fig. \ref{Fig:bifurcation}. 
Indeed, the $k=0$ fixed point is stable for 
$\lambda<\lambda_c$ and unstable for $\lambda>\lambda_c$, while 
the reverse is true for the $k=-1$ fixed point.

\section{Domain walls for $\lambda=0$}
\setcounter{equation}{0}

For $\lambda=0$ the equations (\ref{DS}) become
\be\label{DS0}
\dot u =  - \alpha uv \, ,\qquad 
\dot v =   - \beta v^2 - {1\over 2\alpha} u^2 -2\beta\Lambda\, ,
\ee
and the constraint (\ref{con}) becomes
\be\label{conzero}
\beta^2\left( v^2 -u^2 + 2\Lambda\right) = ke^{-2\beta\varphi}\, .
\ee

We shall first consider the special solutions obtained by setting 
$u\equiv0$. We then obtain the exact phase-plane trajectories for 
all solutions, and present an exact solution corresponding to a 
separatrix trajectory. 

\subsection{Some special solutions}
\label{sec:special}

For $u\equiv0$, the equations (\ref{DS0}) reduce to
\be\label{uzero}
\dot v =   - \beta\left( v^2  +2\Lambda\right)\, .
\ee
For $\Lambda<0$ there are three solutions, with $k=0,-1,1$:

\begin{itemize}

\item $k=0$. This is the fixed point solution with $v^2=2 |\Lambda|$, 
and hence $\varphi = \sqrt{2|\Lambda|}\,  z + \varphi_0$ for constant 
$\varphi_0$, which we may set to zero 
without loss of generality, so the domain wall metric is 
\be\label{adSMink}
ds^2_d = dz^2 + e^{2\beta\sqrt{2|\Lambda|}\,  z} ds^2_{d-1}(Mink)\, . 
\ee
where ``{\it Mink}''  indicates a Minkowski metric. This is just 
$adS_d$ foliated by Minkowski hypersurfaces. For standard Minkowski
coordinates, this yields $adS_d$ in horospherical coordinates. 

\item $k=-1$. In this case $|v|< \sqrt{2|\Lambda|}$ and
\be
v= \sqrt{2|\Lambda|} \tanh\left(\beta\sqrt{2|\Lambda|}\, z\right)\, .
\ee
The constraint (\ref{conzero}) becomes
\be
e^{-2\beta\varphi} = \beta^2\left(2|\Lambda| -v^2\right) = 
{2\beta^2 |\Lambda| \over 
\cosh^2\left(\beta\sqrt{2|\Lambda|}\, z\right)}\, ,
\ee
and hence the metric is
\be
ds^2_d = dz^2 + {1\over 2\beta^2 |\Lambda|} \cosh^2 
\left( \beta \sqrt{2|\Lambda|}\, z \right) ds^2_{d-1}(adS)\, . 
\ee
This is $adS_d$ foliated by anti-de Sitter hypersurfaces; 
the $d=5$ case is well-known \cite{Karch:2000ct,Bak:2003jk}.

\item $k=1$. In this case $|v|> \sqrt{2|\Lambda|}$ and
\be
v = \sqrt{2|\Lambda|} \coth\left(\beta\sqrt{2|\Lambda|}\, z\right)\,. 
\ee
The constraint (\ref{conzero}) becomes
\be
e^{-2\beta\varphi} = \beta^2\left(v^2 - 2|\Lambda|\right) = 
{2\beta^2 |\Lambda| \over 
\sinh^2\left(\beta\sqrt{2|\Lambda| z}\right)}\, ,
\ee
and hence the metric is
\be
ds^2_d = dz^2 + {1\over 2\beta^2 |\Lambda|} \sinh^2 
\left( \beta \sqrt{2|\Lambda|} z \right) ds^2_{d-1}(dS)\, . 
\ee
This is $adS_d$ foliated by de Sitter hypersurfaces  
\cite{DeWolfe:1999cp,LopesCardoso:2001rt}.

\end{itemize}
\noindent
Thus, just as de Sitter space can be viewed as an FLRW cosmology for 
which spatial sections can have zero, positive or negative  curvature, 
so anti-de Sitter space can be viewed as a domain wall 
spacetime for which the  wall has zero, positive or negative
curvature. In other words, $d$-dimensional anti de Sitter space can 
be foliated by $(d-1)$-dimensional leaves with Minkowski ($k=0$), 
de Sitter ($k=1$) or anti-de Sitter ($k=-1$) geometry. 

Continuing with this analogy, we observe that since anti de Sitter 
space can be viewed as an FLRW cosmology with $k=-1$ (but not for 
$k=0$ or $k=1$) we would expect de Sitter space to appear as a domain 
wall for $k=1$ (but not for $k=0$ or $k=-1$). Indeed, for $\Lambda>0$ 
there is a  solution of (\ref{uzero}) only if $k=1$, and this solution is
\be
v= \sqrt{2\Lambda} \cot \left(\beta \sqrt{2\Lambda}\, z\right)\, . 
\ee
The constraint (\ref{conzero}) is now
\be
e^{-2\beta\varphi} = \beta^2\left(v^2 + 2\Lambda\right) = 
{2\beta^2\Lambda \over \sin^2\left(\beta\sqrt{2\Lambda}\, z\right)}\, ,
\ee
and hence the metric is
\be\label{desitterdesitter}
ds^2_d = dz^2 +  {1\over 2\beta^2 |\Lambda|}\, 
\sin^2\left(\beta\sqrt{2\Lambda}\, z\right) 
ds^2_{d-1}(dS)\, . 
\ee
This is de Sitter space foliated by de Sitter 
hypersurfaces  \cite{Alishahiha:2004md}. 

Finally, returning to $\Lambda<0$, we consider the $k=-1$ fixed 
point at $v=0$ and $u=\lambda_c \sqrt{|\Lambda|}$.  This has 
$\varphi=\varphi_0$ for constant $\varphi_0$, which is determined 
by the constraint (\ref{conzero}) to be such that
\be
e^{-2\beta\varphi_0} =  {|\Lambda| \over (d-2)^2}\, . 
\ee
The metric is therefore
\be
ds^2_d = dz^2 + {(d-2)^2\over |\Lambda|} \, ds^2_{d-1}(adS)\, . 
\ee
This is a cylindrical spacetime with a $(d-1)$-dimensional anti-de 
Sitter cross-section:
It is the domain wall analog of the Einstein Static Universe.  
It is also the ``curious linear dilaton'' solution found in 
 \cite{Freedman:2003ax}.

\subsection{Exact phase-plane trajectories}
\label{subsection}

For either sign of $\Lambda$ the phase space trajectories may be found 
exactly by the method used in 
\cite{Townsend:2004zp} to find the cosmological trajectories for 
$\Lambda<0$. From (\ref{DS0}) it follows that
\be
\left(\beta v^2 +{1\over 2\alpha} u^2 +
2\beta\Lambda\right)du -\alpha uv \, dv=0
\ee
on any trajectory. The left hand side is not an exact differential 
but if $u>0$ then the function $u^{-(d+1)/(d-1)}$ is an integrating 
factor and this leads to the conclusion that 
\be\label{curve}
v^2 -u^2 + 2\Lambda = - cu^{2/(d-1)}\, ,
\ee
for some constant $c$. A sketch of the phase plane shows that there 
are no trajectories on which $u$ changes sign, and also that all 
trajectories with $u<0$ are mirror images of those with $u>0$, so 
we may restrict the discussion to follow to $u>0$. From a comparison 
of (\ref{curve}) with the 
constraint (\ref{con}) we learn that
\be
cu^{2/(d-1)} = -{k\over  \beta^2}e^{-2\beta\varphi}\, .
\ee
This determines the value of $\varphi$ at any given 
point on a trajectory specified by the 
constant $c$, except on the $k=0$ trajectories, which are obtained 
by the choice $c=0$, and the trajectories with $u\equiv0$, which 
correspond to $|c|=\infty$. Note that 
\be
{\rm sign}\, c = -k\, .
\ee

{}For $\Lambda>0$ the interpretation of (\ref{curve}) is straightforward. 
Each trajectory with $u>0$ corresponds to one choice of $c$, with
$k=1$ for $c<0$ and $k=-1$ for $c>0$. The phase-plane plot is shown in 
Fig. 2b. 

{}For $\Lambda<0$ the interpretation of (\ref{curve}) is not so
straightforward because of the fixed points, as shown in the 
phase-plane plot in Fig 2a.  
Observe that  (\ref{curve}) is solved for any $c$ by  
$(u,v)=(0,\pm\sqrt{2|\Lambda|})$, which are the $k=0$ fixed points,  
but a sketch of the phase plane shows that there are $k=-1$  
trajectories that do not have any fixed point as a limit point. 
The resolution of this puzzle is that a solution of (\ref{curve}) 
for given $c$ may have more than one branch; in other words, each 
value of $c$ may yield more than one trajectory.  For $c<0$ this 
is trivially true because no such trajectory passes through $v=0$; 
each trajectory with $v>0$ therefore has a mirror image with $v<0$.  
The same is true for $c>0$ provided
$c<\bar c$, where
\be
\bar c = (d-1)\left({2|\Lambda|\over d-2}\right)^{(d-2)/(d-1)}\, . 
\ee
In such cases we may restrict attention to the quadrant of the phase 
plane with $u<0$ and $v>0$, in which the curve (\ref{curve}) specifies 
a unique trajectory for given $c<\bar c$. In contrast, the $c>\bar c$ 
trajectories  pass through $v=0$, and for these one must allow for 
both positive and negative $v$. For a given value of $c>\bar c$, the 
curve (\ref{curve}) has two branches in the $u>0$ half-plane.
On one branch the trajectory is asymptotic to both branches of the 
$k=0$ hyperbola. On the other branch, the trajectory has limit points 
at the $k=0$ fixed points. As these fixed points correspond to 
$adS_d$ spacetimes, the interpolating trajectories correspond 
to solutions that are asymptotic to $adS_d$ in either of two
directions. These ``two-faced'' solutions were called ``Janus'' solutions
in \cite{Freedman:2003ax} (by analogy with the Janus solution of IIB
supergravity \cite{Bak:2003jk}).

For $c= \bar c$, and $u>0$, (\ref{curve}) describes the four separatrices that 
meet at the $v=0$ fixed point. One of these interpolates between this fixed point 
and the  $u=0$ fixed point with $v>0$. This separatrix trajectory is therefore 
one of the curves described by the equation
\be
v^2 = u^2  + 2|\Lambda| - (d-1)\left(\lambda_c^2|\Lambda| \right)^{(d-2)/(d-1)}\, 
u^{2/(d-1)}\, . 
\ee

\subsection{An exact separatrix solution}
\label{sec:exact}

Let us consider in more detail the $d=3$ case, for which we may 
write (\ref{curve}) as
\be\label{quadratic}
v^2 -\left(u - c/2\right)^2 = -{1\over4}\left(c^2 + 8\Lambda\right)\, . 
\ee
For $\Lambda<0$ there is clearly a change of behaviour of the trajectories when $c^2= 8|\Lambda|$, and for $u>0$ this occurs when $c=\sqrt{8|\Lambda|} \equiv \bar c$. At this critical 
value of $c$, the equation (\ref{quadratic} ) degenerates to 
\be
v^2 = \left(u- \sqrt{2|\Lambda|}\right)^2\, . 
\ee

\begin{figure}[h]
  \vskip1em
  \begin{center}
    (a)\epsfig{file=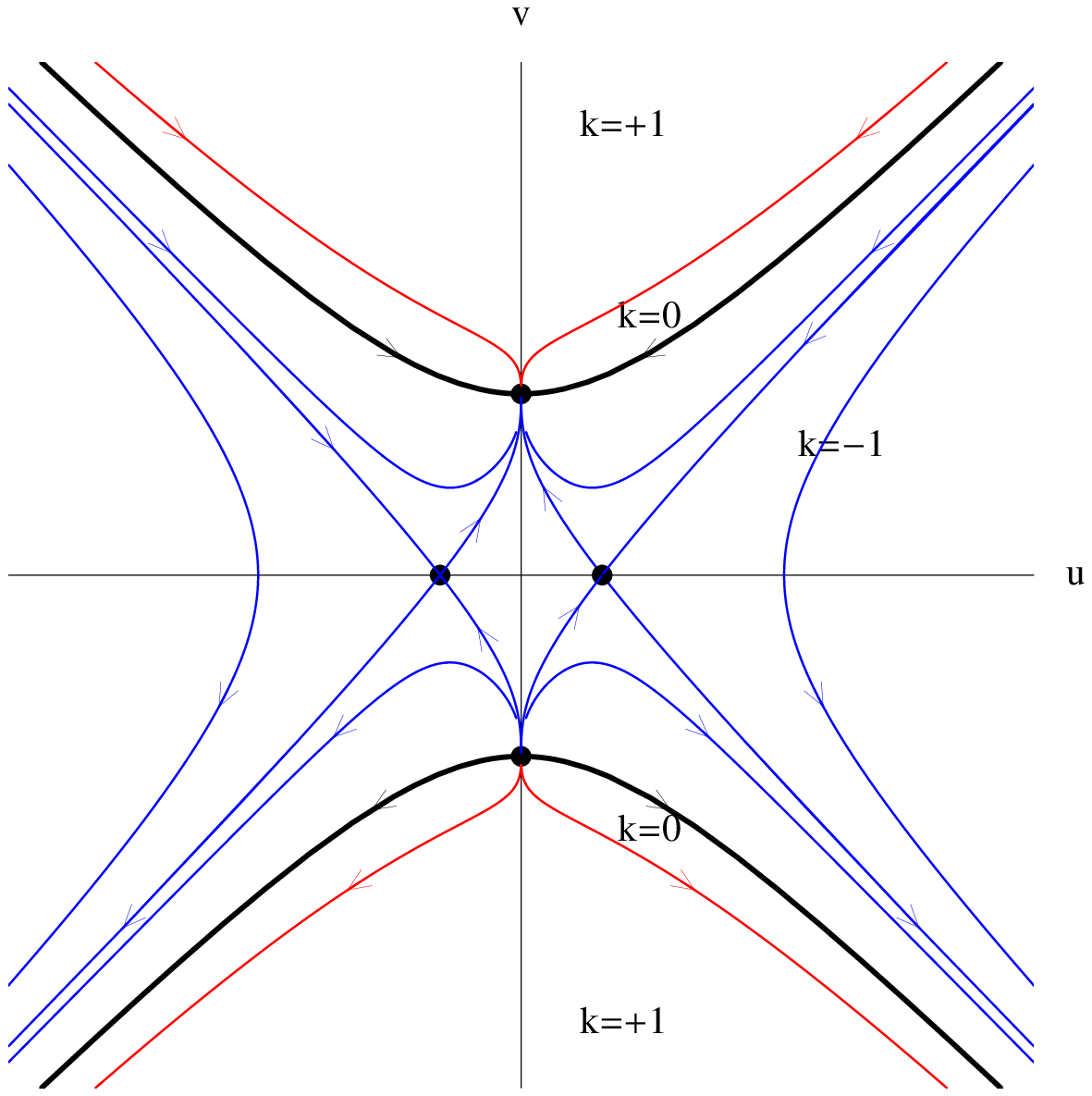,width=6.5cm}\hskip2em(b)\epsfig{file=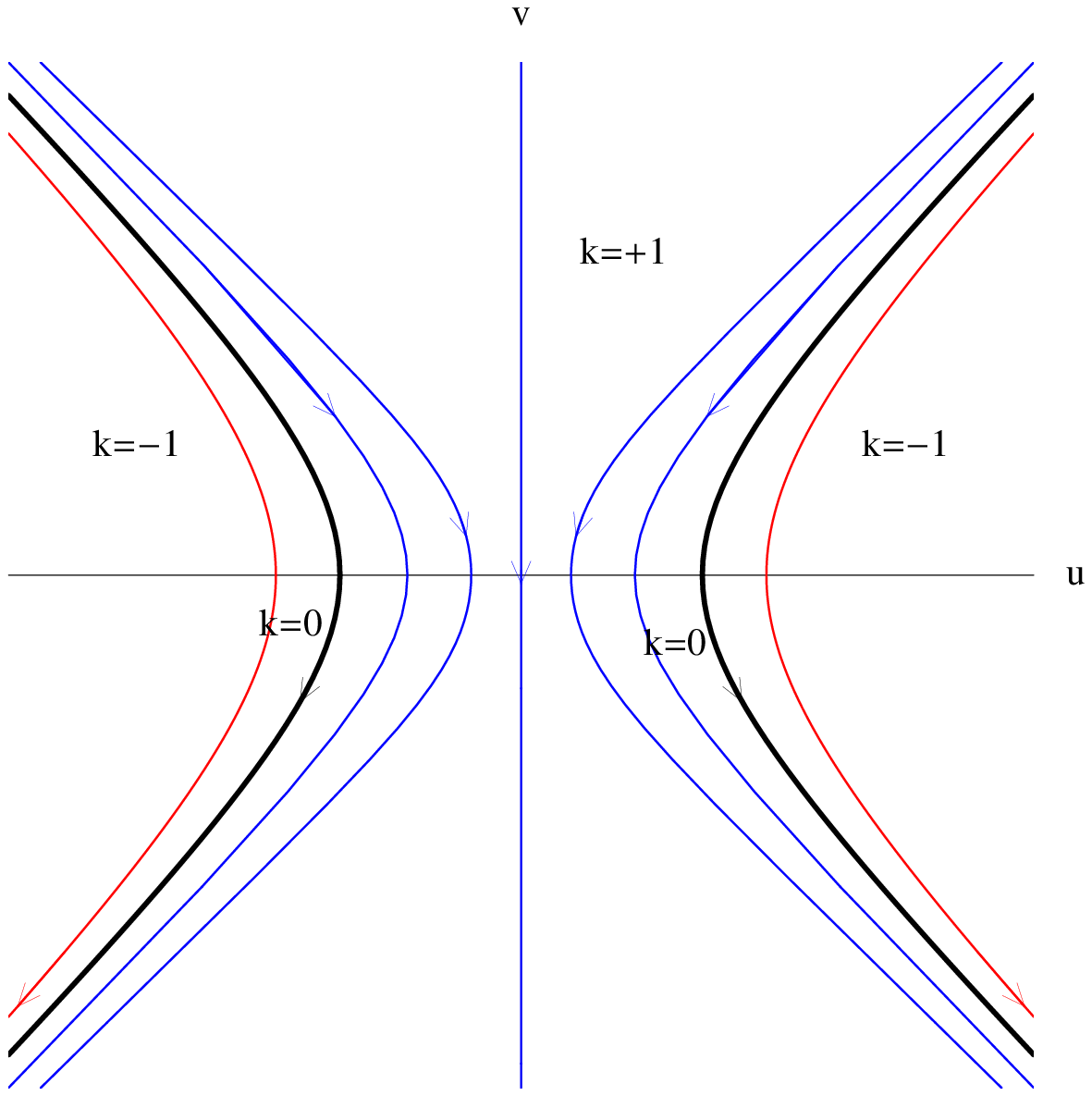,width=6.5cm}
    \begin{picture}(2,2)(0,0)
      \put(-360,190){$\Lambda<0$,\, $\lambda =0$,\,}
      \put(-138,190){$\Lambda>0$,\, $\lambda =0,$\,}
      \end{picture}
     \end{center}\caption{\small{(a) The phase plane for
      $\Lambda<0$. There are four fixed points, connected by separatrices.
  The solutions corresponding to trajectories along the $v-$axis are
  foliations of $adS$. (b) The phase plane for $\Lambda>0$. The straight 
line trajectory
      along the $v-$axis corresponds to the $dS$ foliation of de Sitter space
      (\ref{desitterdesitter}).}}\label{zerodiagram}
\end{figure}

This describes four straight-line separatrices that  meet at the 
$(u,v)=(\sqrt{2|\Lambda|},0)$ fixed point. In particular the
separatrix that interpolates between this fixed point and the 
$(u,v)=(0, \sqrt{2|\Lambda|})$  fixed point is the straight line
\be\label{straightline}
u+v =  \sqrt{2|\Lambda|}\, . 
\ee
On this line, the first of equations (\ref{DS0}) becomes
\be
\dot u = - u\left(\sqrt{2|\Lambda|} -u\right)\, . 
\ee
This equation is easily integrated; taking into account that 
$u<\sqrt{2|\Lambda|}$ on the separatrix, we find that
\be
(u,v) = {\sqrt{2|\Lambda|}\over e^{\sqrt{2|\Lambda|}\, z} +1} 
\, \left( 1,\, e^{\sqrt{2|\Lambda|}\, z} \right)\, .
\ee
As $k=-1$ on the separatrix, the constraint (\ref{conzero}) 
implies that
\be
e^\varphi = {4\over u^2-v^2 + 2|\Lambda|} = |\Lambda|^{-1} 
\left(1+ e^{\sqrt{2|\Lambda|}\, z} \right)\, ,
\ee
and hence that the metric is 
\be
ds^2_d =dz^2 + |\Lambda|^{-1} \left(1+ e^{\sqrt{2|\Lambda|}\, z} 
\right) d\Sigma_{-1}^2 \, . 
\ee
To complete the solution, we observe that (\ref{straightline}) 
implies  $\dot\sigma = \sqrt{2|\Lambda|} - \dot\varphi$, and hence that
\be
e^{\sigma-\sigma_0} = \left[1+ e^{-\sqrt{2|\Lambda|}\, z} \right]^{-1} 
\ee
for some constant $\sigma_0$. 

Thus, for $d=3$, we have found the {\it exact} separatrix solution, 
and not merely the exact phase-plane trajectory. As $z\to -\infty$ we have
\be
\varphi \sim -\log |\Lambda| \, , \qquad
\sigma \sim \sigma_0 + \sqrt{2|\Lambda|} \, z\, , 
\ee
which yields the $adS_2\times \bR$ solution at the $k=-1$ fixed
point. 
As $z\to \infty$ we have 
\be
\varphi \sim \sqrt{2|\Lambda|}\, z -\log |\Lambda|\, ,
\qquad \sigma \sim \sigma_0\, , 
\ee
which yields the $adS_3$ solution at the $k=0$ fixed point. 
\section{Domain walls for $\lambda>0$}
\setcounter{equation}{0}
\begin{figure}[ht]
\vskip1em
  \begin{center}
    (a)\epsfig{file=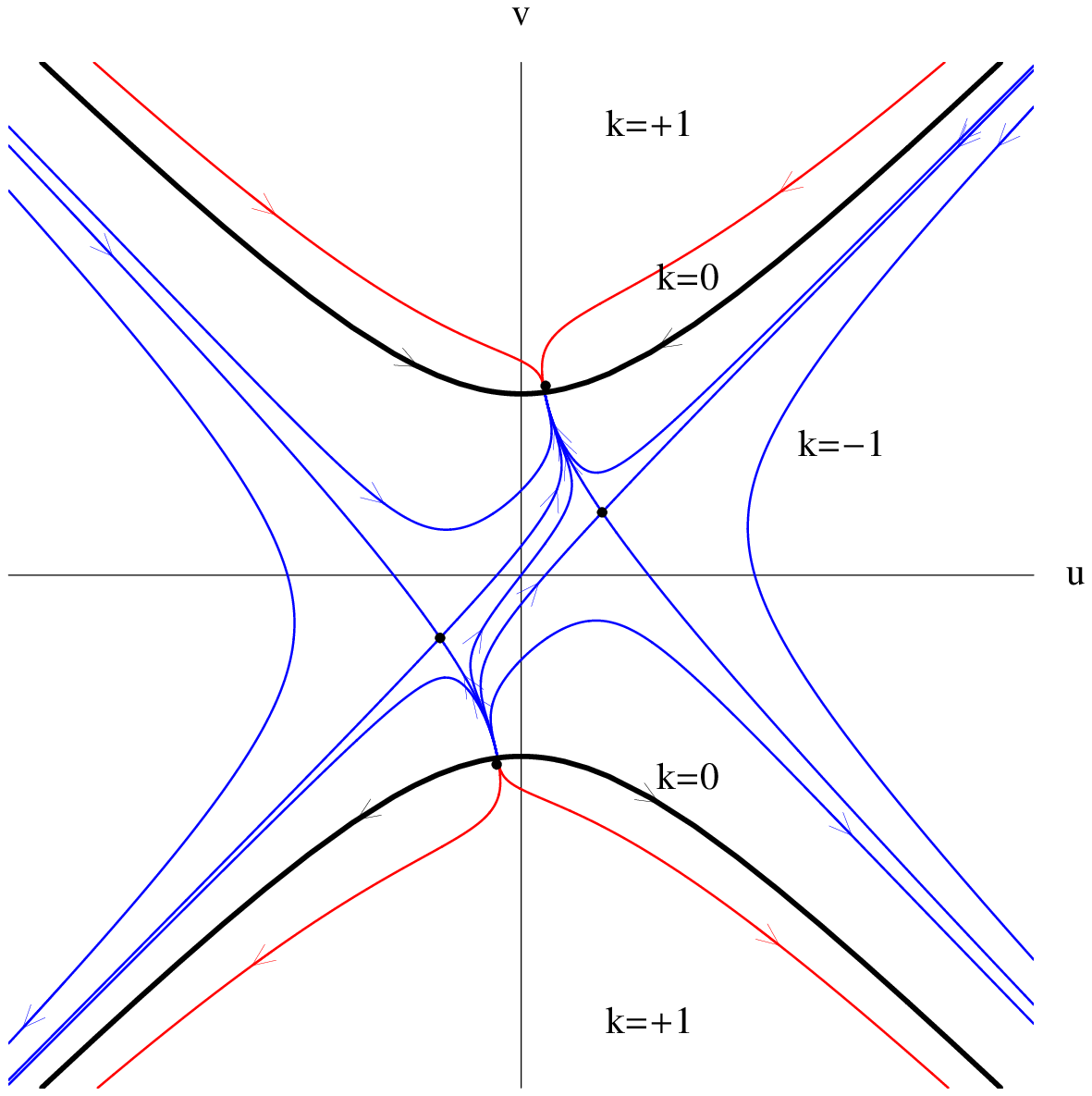,width=6.5cm}\hskip2em(b)
\epsfig{file=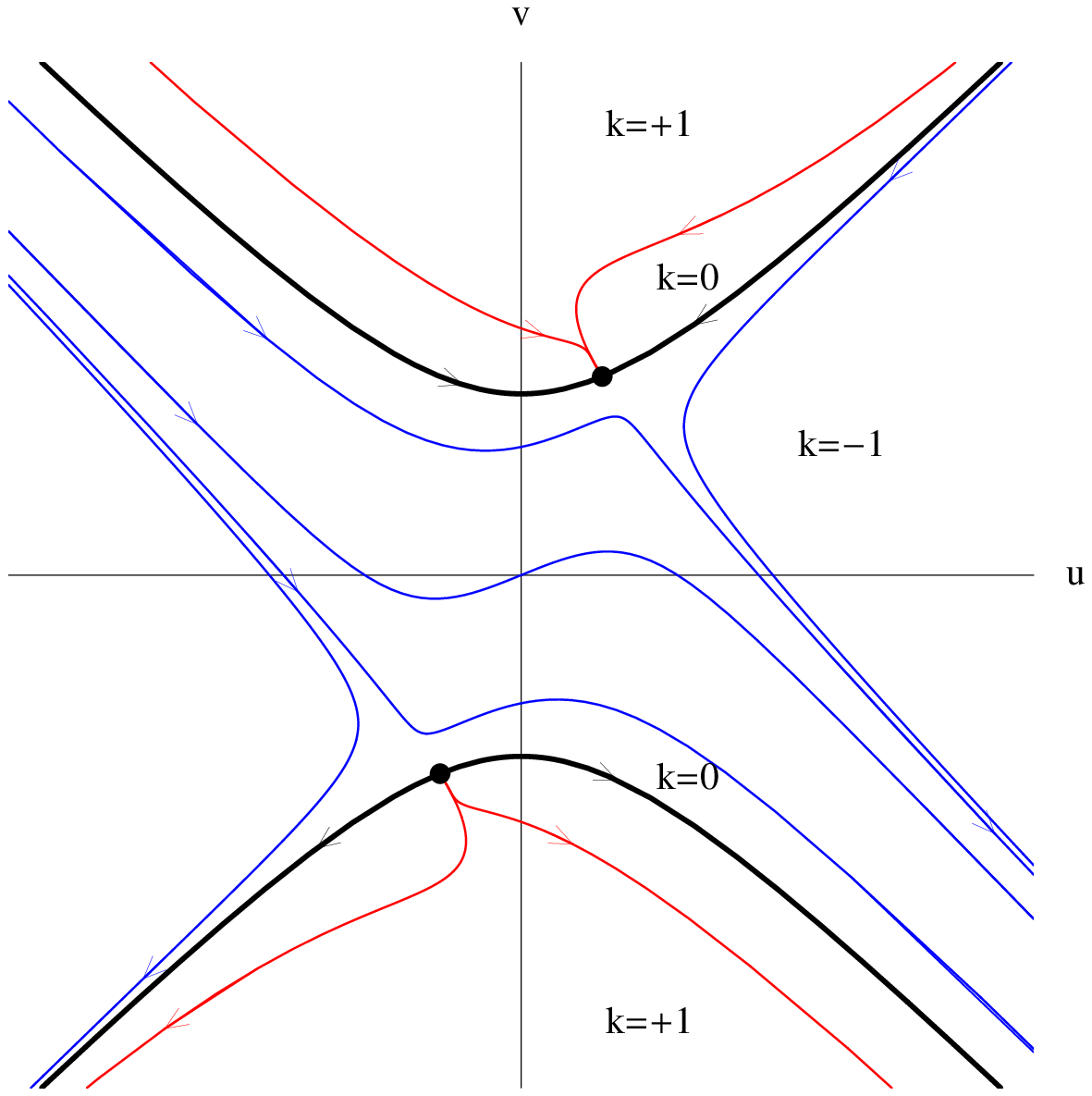,width=6.5cm}
    \begin{picture}(2,2)(0,0)
      \put(-140,210){$\Lambda<0$,\, $\lambda <\lambda_c$}
      \put(82,210){$\Lambda<0$,\, $\lambda =\lambda_c$}
      \end{picture}
     \end{center}\caption{\small{(a) The
      phase plane has the same topology as for $\lambda=0$, with four
      hyperbolic fixed points, but the lateral symmetry is lost. 
(b) The four fixed points 
have coalesced to form a
      pair of non-hyperbolic fixed points.}}\label{nonzerodiagram}
\end{figure}
For $\lambda>0$ our first task is to identify the nature of the domain 
wall spacetimes corresponding to the fixed points.  We then find
exactly  {\it all} $k=0$ solutions, following the method used in 
\cite{Townsend:2003qv} for cosmology.  For generic $k=0$ trajectories 
we fall back on a qualitative analysis of the phase-plane; see 
Figs. 3,4,5.

\subsection{Fixed point solutions}
\label{sec:fixed}

\begin{figure}[t]
  \begin{center}
    (a)\epsfig{file=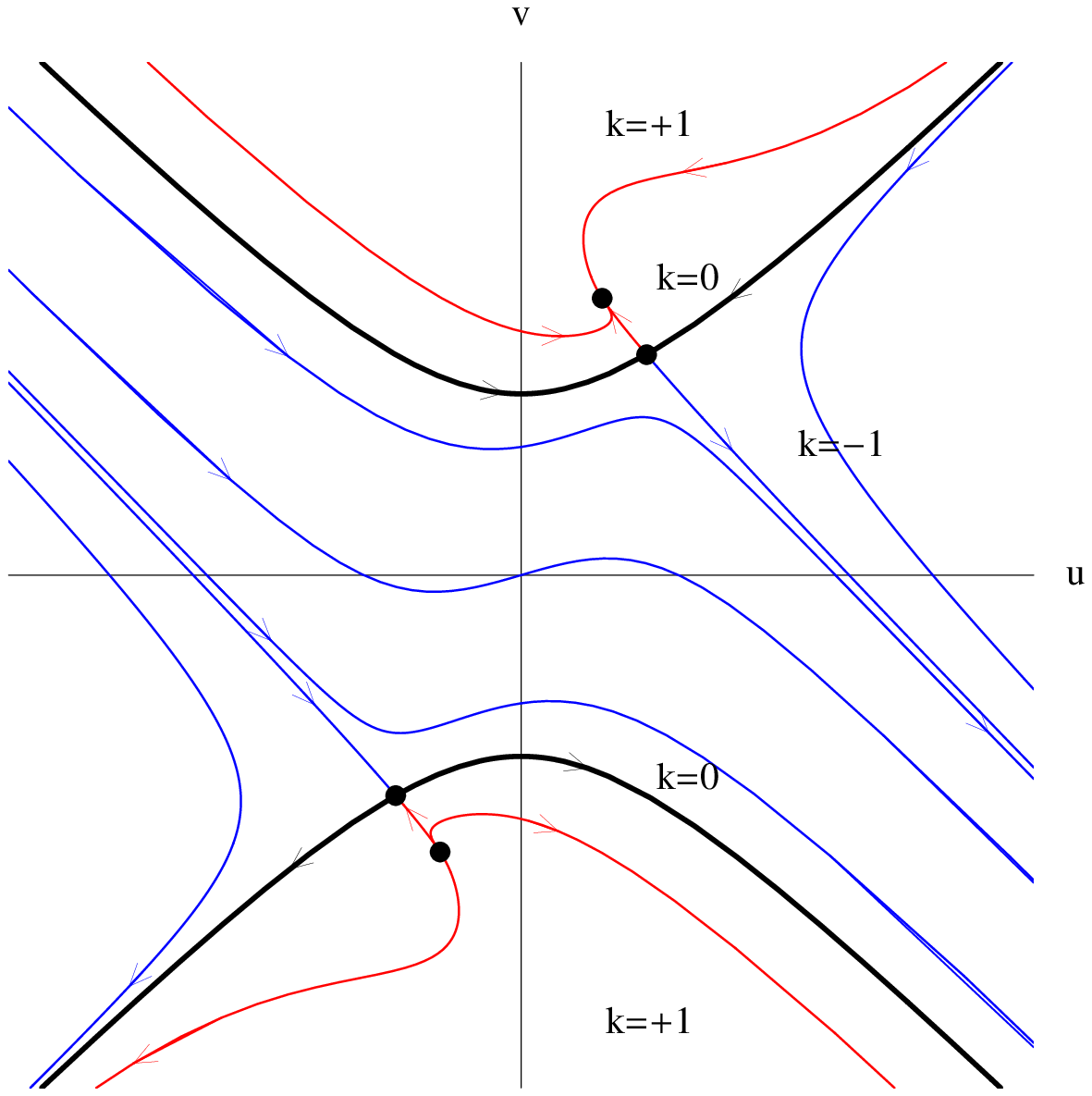,width=6.5cm}\hskip2em(b)
\epsfig{file=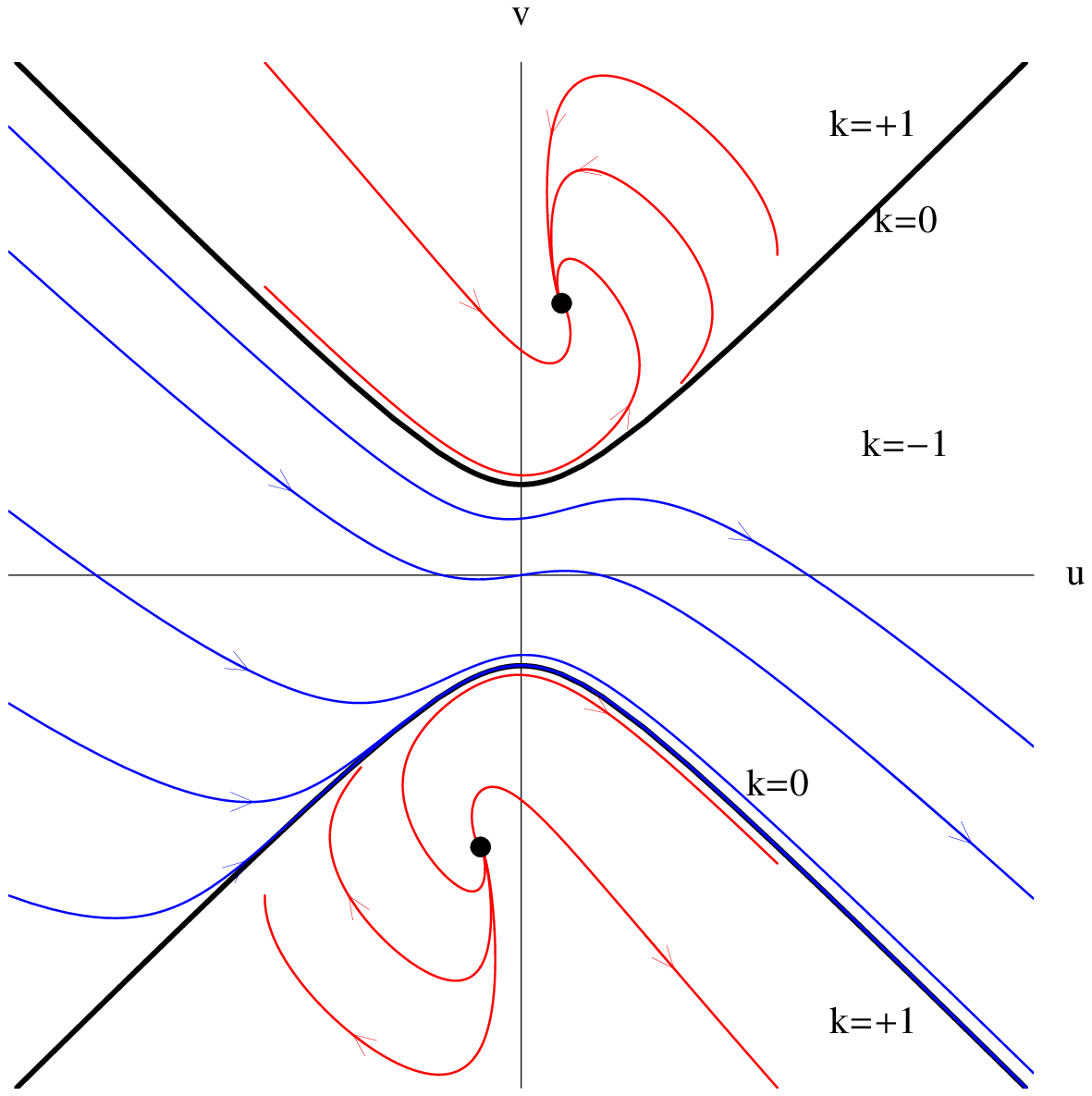,width=6.5cm}
    \begin{picture}(2,2)(0,0)
      \put(-155,205){$\Lambda<0$,\, $\lambda_c<\lambda<2\alpha$}
      \put(80,205){$\Lambda<0$,\, $\lambda >2\alpha$}
      \end{picture}
     \end{center}\caption{\small{(a) The two $k\ne0$ fixed points 
are now in the $k=-1$ regions.  (b) The $k=0$ fixed points have
      disappeared to $\infty$.}}\label{nonzerodiagram2}
\end{figure}

We need consider only the fixed points with $u>0$. We consider each of 
the types of fixed point in turn. 

\begin{itemize}

\item Type 1. At this fixed point we have
\be
\sigma = \lambda K(\lambda) z + \sigma_0 \, , \qquad
2\beta \varphi = \lambda_c^2 K(\lambda) z + 2\beta\varphi_0
\ee
for constants $\sigma_0$ and $\varphi_0$. Without loss of generality, 
we may choose
\be
\sigma_0 = {2\over\lambda} \log\left(\lambda^2K/2\right)\, ,\qquad
\varphi_0=0\, ,
\ee
in which case the  fixed-point solution is
\be
ds^2_d =  dr^2 + r^{2\lambda_c^2/\lambda^2}ds^2_{d-1}(Mink)\, , \qquad
e^{{1\over2}\lambda\sigma} = {1\over2} \lambda^2 K \, r \, , 
\ee
where 
\be
r^2= e^{\lambda^2K(\lambda) z}\, . 
\ee
This solution was first found in  \cite{Lu:1995hm}.
Recall that $\lambda<2\alpha$ when $\Lambda<0$; in particular, there 
is a fixed point for $\lambda=\lambda_c$  when $\Lambda<0$, with 
fixed-point solution
\be\label{lambdac}
ds^2_d = dr^2 + r^2ds^2_{d-1}(Mink)\, ,\qquad
e^{{1\over2}\lambda_c\sigma} ={\sqrt{|\Lambda|}\over (d-2)} \, r\, . 
\ee

\item Type 2. In this case $\Lambda<0$, necessarily. At this fixed point we have
\be
\sigma = \lambda_c \sqrt{|\Lambda|}\, z + \sigma_0 \, ,\qquad
2\beta \varphi = \lambda \lambda_c \sqrt{ |\Lambda|}\,  z + 2\beta\varphi_0 
\ee
for constants $\sigma_0$ and $\varphi_0$. We may assume that 
$\lambda\ne\lambda_c$ because the coincidence of the fixed points at
$\lambda=\lambda_c$ means that the fixed point solution is the same as the 
$\lambda=\lambda_c$ case of the Type 1 solution discussed above. For 
$\lambda\ne\lambda_c$ we are {\it not} free to choose the constants 
$(\sigma_0,\varphi_0)$ arbitrarily because the constraint (\ref{con}) requires 
\be
\lambda\sigma_0 -2\beta\varphi_0 = 
\log\left[{\left|\left(\lambda^2-\lambda_c^2\right)\Lambda\right| 
\over 2(d-2)}\right]\, . 
\ee
However, we may choose
\be
\sigma_0 = {2\over\lambda} \log \left(\lambda \lambda_c 
\sqrt{|\Lambda|} /2\right)\, ,
\ee
without loss of generality, in which case the fixed-point solution is
\be
ds^2_d = d\rho^2 + {\lambda^2 \over |\lambda^2-\lambda_c^2|}\, 
\rho^2 d\Sigma_k^2 \, , \qquad 
e^{{1\over2}\lambda\sigma} = {1\over2} \lambda \lambda_c 
\sqrt{|\Lambda|}\,  \rho \, ,
\ee
where
\be
\rho^2 = e^{\lambda  \lambda_c \sqrt{|\Lambda|}\,  z}\, . 
\ee

For $\lambda=0$, the fixed point is in the region of the phase plane 
with $k=-1$ and we thus recover the 
$adS_{d-1}\times \bR$ product metric found in the previous section. 
For $\lambda>\lambda_c$ the fixed point is in the $k=1$ region of 
the phase plane and the fixed point solution is
\be
ds^2_d = d\rho^2 + {\lambda^2 \over |\lambda^2-\lambda_c^2|}\, 
\rho^2 ds^2_{d-1}(dS) \, , \qquad 
e^{{1\over2}\lambda\sigma} = {1\over2} \lambda \lambda_c 
\sqrt{|\Lambda|}\,  \rho \, .
\ee
In the limit as $\lambda\to\infty$ the $d$-metric becomes a flat static 
Rindler-type metric that is the analytic continuation of the Milne metric 
through its cosmological horizon; this is possible because
the stress tensor for $\sigma$ is proportional to $1/\lambda^2$. 

\end{itemize}

\subsection{Flat walls}
\label{sec:flat}

Flat domain walls, for which the worldvolume geometry is Minkowski, 
are found by considering $k=0$. 
In this special case, the equations (\ref{DS}) reduce to 
\be\label{ueq}
\dot u = {1\over2}\left(\lambda v - 2\alpha u\right) v\, , \qquad
\dot v = {1\over2}\left(\lambda v -2\alpha u\right) u\, , 
\ee
and the constraint is  
\be\label{flatcon}
v^2-u^2 + 2\Lambda =0\, . 
\ee
Solving the constraint by setting
\be
v= \sqrt{|\Lambda|/2}\left(\xi - {\rm sign}\Lambda\, \xi^{-1}\right)\, ,\qquad
u= \sqrt{|\Lambda|/2}\left(\xi + {\rm sign}\Lambda\, \xi^{-1}\right)\, ,
\ee
we find that the equations (\ref{ueq}) are equivalent to 
\be\label{xiequation}
\dot\xi = {1\over4}\sqrt{2|\Lambda|}\left[
  \left(\lambda-2\alpha\right)\xi^2 
-  ({\rm sign}\,\Lambda)\, \left(\lambda+2\alpha\right)\right]\, .
\ee

For future convenience, we choose to present the solutions for
$\lambda=0$ 
and $\lambda>0$ separately.

\subsubsection{$\lambda=0$}
In this case (\ref{xiequation}) reduces to
\be
\dot\xi = -\alpha\sqrt{|\Lambda/2}\left({\rm sign}\, 
\Lambda + \xi^2\right)\, .
\ee
This is easily solved and leads to the following solutions for 
$(\varphi,\sigma)$:
\begin{itemize}

\item $\Lambda>0$. In this case
\be\label{lzero2}
e^{\alpha\varphi} = {1\over2}  \left|\sin \left(\alpha 
\sqrt{2\Lambda}\, z\right)\right|\, ,\qquad
e^{\alpha\sigma } = \left|\cot \left(\alpha\sqrt{\Lambda/2}\, 
z \right)\right|\, . 
\ee
Formally, this is a periodic solution with period 
$\pi/ [\alpha \sqrt{2\Lambda}]$ but because
$\varphi$ is singular at $z=0$ we should consider $z>0$ and 
$z<0$ as yielding different solutions. Moreover we may restrict 
to $\alpha \sqrt{2\Lambda} \ |z| <\pi$ as $\sigma$ is singular when
$\alpha \sqrt{2\Lambda} \ |z| =\pi$. The two solutions with $z>0$ 
and $z<0$ yield the solutions corresponding to the two branches of 
the $k=0$ hyperbola. 

\item $\Lambda<0$. In this case there is a fixed point solution, 
which we have already discussed. Otherwise, we have
\be\label{lzero1}
e^{\alpha\varphi }= {1\over2} \left| \sinh 
\left(\alpha \sqrt{2|\Lambda|}\ z\right) \right|\,, \qquad
e^{\pm\alpha\sigma} = \left| 
\coth \left(\alpha \sqrt{|\Lambda|/2}\, z\right) \right|\,  . 
\ee
For either choice of the sign we have two solutions, corresponding to 
$z>0$ and $z<0$, since $\varphi$ is singular at $z=0$. These are the 
two branches of the $k=0$ hyperbola, with $\dot\varphi>0$ for $z>0$ 
and $\dot\varphi<0$ for $z<0$. On each branch there are two solutions, 
apart from the fixed point solution, depending on whether $\dot\sigma$ 
is positive or negative; this corresponds to the choice of sign 
in (\ref{lzero1}).

\end{itemize}

\subsubsection{$\lambda>0$}

It is convenient to introduce the quantities
\be
A = {\sqrt{|(4\alpha^2-\lambda^2)\Lambda|} \over 2\sqrt{2}}\, ,\qquad
\nu_\pm = {2\over 2\alpha \pm \lambda}\, . 
\ee
The solutions  can be given jointly for either sign of $\lambda$,  
according to whether $(\lambda-2\alpha)\Lambda$ is positive, negative or zero:

\begin{itemize}
\item $\lambda=2\alpha$. In this case, 
$\xi= -\alpha ({\rm sign}\,\Lambda)\, \sqrt{2|\Lambda|}\, z$, and
\be
e^{2\alpha\varphi} =  z e^{-\Lambda \alpha^2 z^2}\, ,\qquad 
e^{2\alpha\sigma} = z^{-1} e^{-\Lambda \alpha^2 z^2}\, . 
\ee

\item $(\lambda -2\alpha)\Lambda <0$. In this case,
\be
\xi = -({\rm sign}\,\Lambda)\, 
\sqrt{\lambda + 2\alpha  \over |\lambda - 2\alpha|}\, \tan Az\, , 
\ee
and the solution is
\be
e^\varphi = \left|\cos Az\right|^{\nu_- } 
\left|\sin Az\right|^{\nu_+} \, , \qquad
e^\sigma =  \left|\cos Az\right|^{\nu_- } 
\left|\sin Az\right|^{-\nu_+} \, . 
\ee

\item $(\lambda -2\alpha)\Lambda >0$
In this case, there are two solutions of (\ref{xiequation}), 
in addition to the fixed point solutions already considered:
\bea
(i): \  \xi &=&- ({\rm sign}\,\Lambda)\, 
\sqrt{2\alpha +\lambda \over 2\alpha -\lambda}\, 
\tanh Az\, ,\nonumber\\
(ii):\  \xi &=&  - ({\rm sign}\,\Lambda)\, 
\sqrt{2\alpha +\lambda \over 2\alpha -\lambda}\, \coth Az\, .
\eea
These yield the solutions
\bea
(i)\qquad e^\varphi &=& \left(\cosh Az\right)^{\nu_-} 
\left|\sinh Az\right|^{\nu_+} \, , \nonumber\\
e^\sigma &=& \left(\cosh Az\right)^{\nu_-} 
\left|\sinh Az\right|^{-\nu_+} \, ,
\eea
and 
\bea
(ii)\qquad e^\varphi &=&  \left(\cosh Az\right)^{\nu_+} 
\left|\sinh Az\right|^{\nu_-} \, ,\nonumber\\
e^\sigma &=& \left(\cosh Az\right)^{-\nu_+} 
\left|\sinh Az\right|^{\nu_-} \, . 
\eea

\end{itemize}

\subsection{Generic domain walls}

\begin{figure}[t]
  \begin{center}
    (a)\epsfig{file=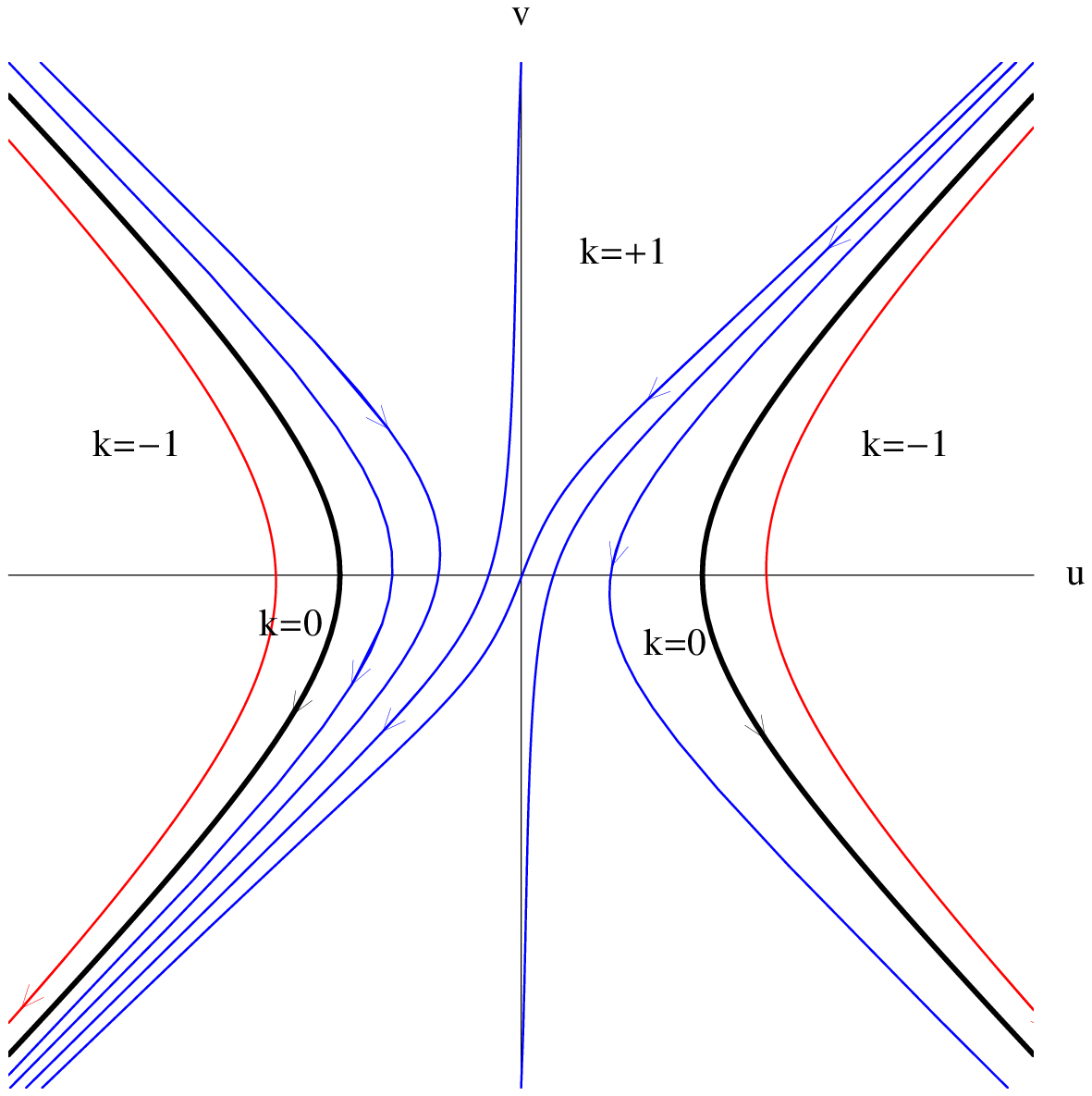,width=6.5cm}\hskip2em(b)
\epsfig{file=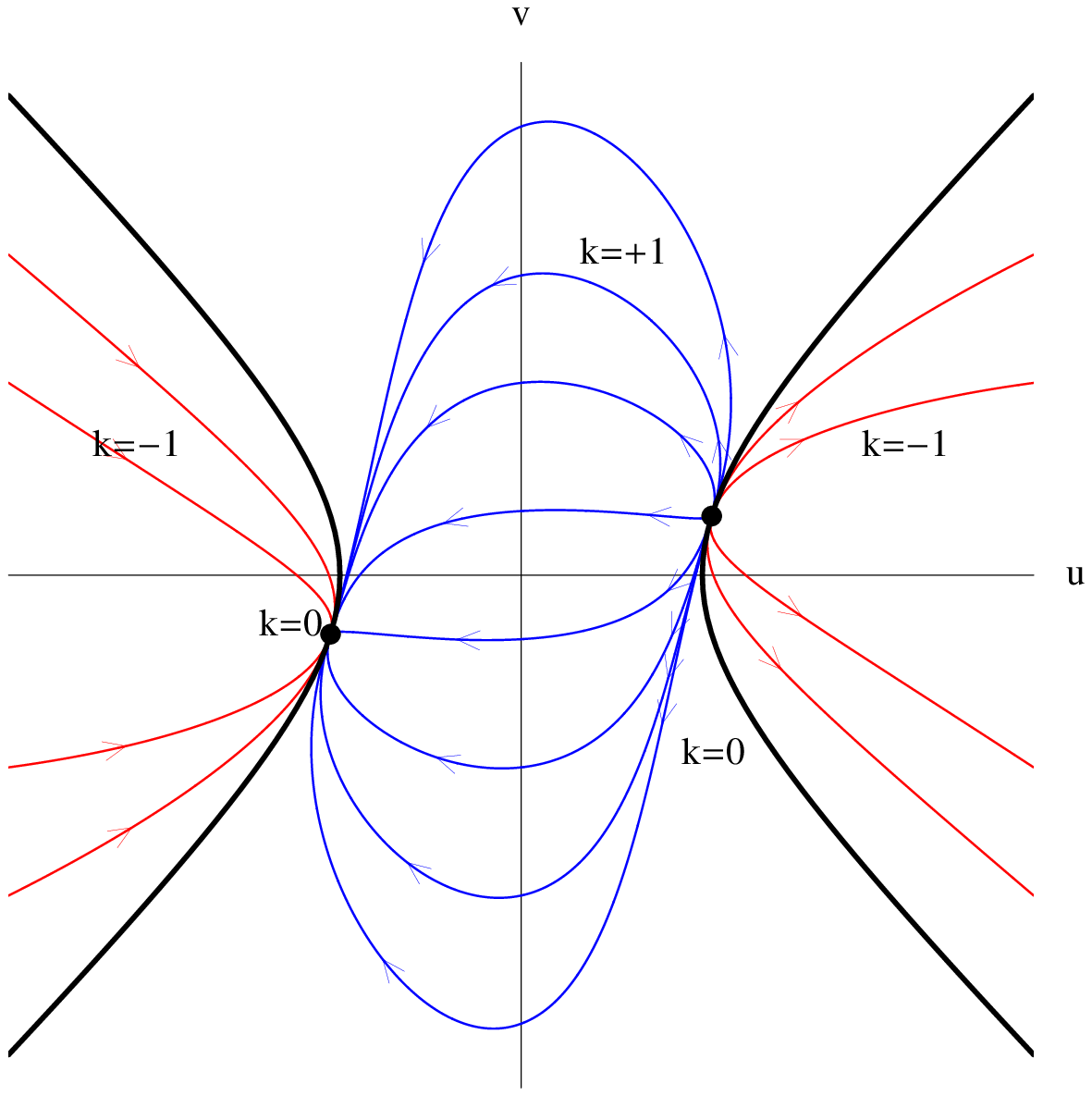,width=6.5cm}
\begin{picture}(2,2)(0,0)
      \put(80,210){$\Lambda>0$,\, $\lambda>2\alpha$}
      \put(-145,210){$\Lambda>0$,\, $0<\lambda <2\alpha$}
      \end{picture}
\end{center}\caption{\small{(a) The phase plane topology is the
      same as for $\lambda=0$ but the lateral symmetry is lost and there
      are now two trajectories that are asymptotic to the $v$-axis. 
(b) For $\lambda>2\alpha$ there are two $k=0$ fixed points, both of
      which are
      nodes. All $k=1$ trajectories start at one node and end at the
      other one.}}\label{nonzerodiagram3}
\end{figure}

A generic trajectory in the $(u,v)$ phase-plane is a solution to the 
differential equation 
\be
\left(\beta v^2 + {1\over 2\alpha^2}u^2 +2\beta\Lambda - 
{1\over2}\lambda uv\right) du
+ \left({1\over2}\lambda u^2 - \alpha uv -
\lambda \Lambda\right) dv = 0\, .
\ee
The left hand side is not an exact differential but an integrating 
factor exists. For $\lambda=0$ we were able to find the integrating 
factor and hence we were able to find all the trajectories exactly. 
We have not found the integrating factor for $\lambda>0$, so in this 
case we must fall back on a qualitative analysis of the phase-plane 
trajectories. However, given that
one has trajectories for any $\lambda$, one can ask what they 
look like in the limit as $\lambda\to \infty$. In this limit,
an integrating factor is easily found and this yields the curves
\be\label{infinity}
Cv^2 -u^2 + 2\Lambda=0
\ee
for some constant $C$, which must be non-negative for $\Lambda<0$ but
may be positive or negative for $\Lambda>0$.

The phase-plane plots are essentially the same as those found 
in the cosmological case  \cite{Halliwell:1986ja}  for $d=4$ (and $\Lambda>0$, 
which corresponds here to $\Lambda<0$) but with the different 
interpretation discussed earlier. As there are are never more than four 
hyperbolic fixed points, the topological structure of the phase-plane is 
determined unambiguously by these fixed points as long as there are 
no limit cycles. For $\Lambda>0$ or for $\Lambda<0$ when 
$\lambda<\lambda_c$, any closed curved in the phase plane with unit 
Poincar\'e index must cross the $k=0$ hyperbola and so cannot be a 
limit cycle. Thus a limit cycle is possible only if $\Lambda<0$ and 
$\lambda>\lambda_c$, and any such cycle would have to enclose
a $k=1$ fixed point (which is  either a node or a focus, as explained below, 
and hence has unit Poincar\'e index).  We have not seen how to prove 
that a limit cycle never appears for any $d$ as $\lambda$ is increased 
indefinitely,  but numerical plots for various cases are consistent with the 
absence of  limit cycles.
We present a representative selection of $\lambda>0$ 
phase-plane plots in Figs. 3,4,5. These were obtained numerically for $d=7$ and
particular choices of $\lambda$ in the specified ranges. Note the
symmetry under reflection through the origin, in all cases. For $\Lambda<0$ and 
$\lambda>2\alpha$ the $k=1$ fixed point is either
a node or a focus (spiral) depending on $d$ and the precise value of $\lambda$.
The details are the same as in the cosmology case
\cite{Townsend:2003qv}. In the 
notation of this paper one finds that the fixed point is a node for
all $d\ge10$ and for $d<10$ if $\lambda_c<\lambda \le \bar\lambda$,  
where\footnote{Note that $\bar\lambda \le 2\alpha$ with equality 
for $d=9$.}
\be
\bar\lambda= {4\over \sqrt{(d-2)(10-d)}}\, . 
\ee
Otherwise, the $k=1$ fixed point is a focus. However, there
is no {\it topological} distinction between a focus and a node. 

Note that for $\Lambda<0$ and $\lambda<\lambda_c$, the phase-plane plot of
Fig. 3a shows that there is a one-parameter family of 
Janus-type solutions that interpolate
between the isometric domain wall spacetimes corresponding to the 
two $k=0$ fixed points. These are deformations of the family of 
Janus solutions of the $\lambda=0$ case.

\section{Supersymmetry}
\setcounter{equation}{0}

For various values of the coupling constant $\lambda$, and choices 
of the sign of $\Lambda$, the Lagrangian density (\ref{origlag}) 
is the consistent truncation of a supergravity Lagrangian density for 
which the metric and dilaton are the only bosonic fields. In 
this context one can ask whether any given solution preserves 
some fraction of the supersymmetry of the supergravity vacuum.  
A necessary condition for (partial) supersymmetry preservation 
is the vanishing of  the dilatino supersymmetry transformation. 
This imposes the condition 
\be\label{susycon}
\left(\Gamma^\mu\partial_\mu \sigma + 2W'\right)\epsilon=0\, ,
\ee
where $\epsilon$ is the supersymmetry spinor parameter, and 
$W(\sigma)$ is the superpotential, which must satisfy
\be\label{superpoteq}
(W')^2 -\alpha^2 W^2 = {1\over2}\Lambda\,  e^{-\lambda\sigma}\, .
\ee
The matrices $\Gamma_\mu$ obey 
the Dirac commutation relations in the given background, which 
is all that we need to know about them, although they may not 
actually be the Dirac matrices\footnote{They
{\it are} the Dirac matrices for $d=3$ minimal supergravity,
but they are not necessarily irreducible for other odd dimensions
(or non-minimal supergravities); for example, 
$\Gamma_\mu= i\sigma_2\otimes \gamma_\mu$ for $d=5$.  
In even dimensions $\Gamma_\mu$ is the product of 
$\gamma_\mu$ with the chirality matrix, in a Majorana basis.}.

Given a metric of the form (\ref{metricchoice}), we may choose
frame 1-forms 
\be
e_z = e^{{1\over2}\lambda \sigma} dz\, , \qquad
e_m = e^{\beta\varphi} \hat e_m\, , 
\ee
where $\hat e_m$ ($m=0,1,\dots,d-2$) are a set of frame 1-forms for
the $(d-1)$ metric $d\Sigma_k^2$ on the wall. In such a frame we have
\be
\Gamma_\mu = (e^{{1\over2}\lambda\sigma}\, \Gamma_z, \ 
e^{\beta\varphi}\, \hat\Gamma_m)\, ,
\ee
where $\Gamma_z$ is a constant matrix that squares to the identity and
anticommutes with the matrices $\hat\Gamma_m$. Given that $\sigma$ is
a function only of $z$, the condition (\ref{susycon}) now reduces to
\be\label{susycon2}
\dot\sigma = \pm 2e^{{1\over2}\lambda\sigma}\, W'\, . 
\ee
If this is satisfied for $\dot\sigma=0$ (and hence constant 
$\sigma$ such that $W'(\sigma)=0$) then there is no condition 
on $\epsilon$. Otherwise
\be\label{halfsusy}
\left(1 \pm \Gamma_z\right)\epsilon=0\, , 
\ee
which implies, in the absence of any further condition on $\epsilon$,
that 1/2 supersymmetry is preserved. 

We must also take into account the Killing spinor condition
\be\label{killspin}
\left(D_\mu - {1\over 2(d-2)}\, W \Gamma_\mu\right)\epsilon =0\, , 
\ee
which arises, in a supergravity context, from the requirement of
vanishing gravitino variation. This is equivalent to the equations
\bea\label{reducedKilling}
\left[\partial_z -{1\over 2(d-2)}e^{{1\over2}\lambda\sigma} W  
\Gamma_z\right]\epsilon &=& 0 \nonumber\\
\left[\hat D_m + {\beta\over2}e^{\beta\varphi}\hat\Gamma_m 
\left(\dot\varphi e^{-{1\over2}\lambda\sigma} \Gamma_z 
- 2\alpha W\right) \right]\epsilon &=& 0\, , 
\eea
where $\hat D_m$ is the covariant derivative on spinors restricted to
the domain wall, and with respect to the frame 1-forms $\hat e_m$.
The second of these equations has the integrabilty condition
\be\label{phisusy2}
\dot\varphi^2 = e^{\lambda\sigma}\left[ 4\alpha^2 W^2 + 
{k\over \beta^2}\, e^{-2\beta\varphi}\right]\, .
\ee
There is a further joint integrability condition of equations 
(\ref{reducedKilling}). Using (\ref{phisusy2}) and
\be\label{ddphi}
\ddot\varphi = {1\over2}\left(\lambda \dot\varphi -2\alpha
\dot\sigma\right)\dot\sigma - {k\over\beta} e^{\lambda\sigma
  -2\beta\varphi}\, ,
\ee
which follows from (\ref{DS}) and (\ref{con}), this remaining
integrability condition can be reduced to
\be
\dot\sigma\left(\dot\sigma + 2e^{{1\over2}\lambda\sigma} W'
\Gamma_z\right)\epsilon =0\, . 
\ee
This is an identity if $\dot\sigma=0$; otherwise it reduces to
(\ref{susycon2}) with $\epsilon$ constrained by (\ref{halfsusy}).
Moreover, (\ref{phisusy2}) can be derived by combining
(\ref{superpoteq}) with the constraint (\ref{con}) and
eliminating $W'$ from the resulting expression by means of 
(\ref{susycon2}). Thus, for domain-wall solutions of the field
equations, the `dilatino' supersymmetry preserving condition 
(\ref{susycon2}) is the Killing spinor integrability condition.

For $k=0$ we may choose cartesian coordinates for which 
$\hat D_m= \partial_m$. In this special case, a spinor satisfying
(\ref{halfsusy}) will be a function only of $z$, and
subject to no further algebraic constraints, iff
\be\label{phisusy}
\dot\varphi = \mp 2\alpha e^{{1\over2}\lambda\sigma}\, 
W  \qquad \qquad (k=0). 
\ee 
This is of course consistent with the integrability condition 
(\ref{phisusy2}), but also fixes the sign of $\dot\varphi$.
The Killing spinor itself is given by integration of 
\be\label{intKill}
\partial_z \epsilon = \mp {1\over 2(d-2)} 
e^{{1\over2}\lambda\sigma} W \epsilon\, .
\ee

Application of these results to the problem in hand requires 
that we find a superpotential $W$ satisfying (\ref{superpoteq}).
It is instructive to consider first the $\lambda=0$ case.

\subsection{$\lambda=0$}

 In this case, (\ref{superpoteq}) reduces to
\be\label{speqzero}
2\left[(W')^2 -\alpha^2 W^2\right] = \Lambda\, .
\ee
There are three  possible superpotentials, which we consider in turn:

\begin{itemize}

\item $2\alpha W = \sqrt{2|\Lambda|}$. 

This applies for $\Lambda<0$. As $W'=0$ it is clear from 
(\ref{susycon2}) that only domain wall solutions with 
$\sigma=\sigma_0$, for constant $\sigma_0$, can be supersymmetric 
for this superpotential, and that in this case the condition for 
supersymmetry reduces to the Killing spinor conditions (\ref{reducedKilling}). 
Writing $\epsilon = \epsilon^+ + \epsilon^-$, 
where $\Gamma_z\epsilon^\pm = \pm \epsilon^\pm$, we find that these 
conditions become
\be
\partial_z \epsilon^\pm = \pm \beta
\sqrt{|\Lambda|\over2}\epsilon^\pm\, ,
\qquad
\hat D_m \epsilon^\pm = {\beta\over2} e^{\beta\varphi} 
\left(\sqrt{2|\Lambda|} \pm \dot\varphi\right)\hat \Gamma_m
\epsilon^\mp\, . 
\ee 
The first of these equations is solved by 
\be
\epsilon^\pm  = e^{\pm \beta \sqrt{|\Lambda|/2} z}\zeta^\pm\, ,
\ee
for $z$-independent spinor $\zeta^\pm$ satisfying
$\Gamma_z \zeta^\pm = \pm\zeta^\pm$. The remaining equations 
then reduce to
\be\label{hateq}
\hat D_m \zeta^\pm = {1\over2} C_\pm \hat\Gamma_m \zeta^\mp\, , 
\ee
where
\be
C_\pm = \beta \left(\sqrt{2|\Lambda|} \pm \dot\varphi\right)
e^{\beta\varphi \mp \sqrt{2|\Lambda|} z}\, . 
\ee
The integrability conditon for (\ref{hateq}) is 
\be
C_+ C_- = -k \, .
\ee
A rescaling of $\zeta^+$ and $\zeta^-$ rescales $C_+$ and $C_-$, leaving
the product $C_+C_-$ unchanged, so we may choose $C_+ = -k C_-$
without loss of generality. This choice leads to
\be
 \dot\varphi = \mp \sqrt{2|\Lambda|}\,\left( {e^{\mp\beta
     \sqrt{2|\Lambda|} z} + k e^{\pm\beta\sqrt{2|\Lambda|}
     z}\over e^{\mp\beta
     \sqrt{2|\Lambda|} z} - k e^{\pm\beta\sqrt{2|\Lambda|}
     z}}\right)\, . 
\ee
This reproduces the solutions of section \ref{sec:special}
for the three foliations
of $adS_d$. Thus, all supersymmetry is preserved by the $adS_d$ solution
{\it irrespective of how it is foliated}.

\item $2\alpha W= \sqrt{2|\Lambda|}\, \cosh \alpha\sigma$. 

This again applies for $\Lambda<0$ and was considered in 
\cite{Freedman:2003ax}. From (\ref{susycon2}) we see 
that only domain wall solutions with
\be\label{susycon3}
\dot\sigma =  \pm \sqrt{2|\Lambda|}\, \sinh \alpha\sigma
\ee
can preserve some fraction of supersymmetry. One possibility is 
$\sigma=0$, in which case we again have $adS_d$ for any $k$ and 
all supersymmetries are preserved. This case is analogous to the 
$adS_d$ solution allowed for constant $W$ but 
with the difference that 
supersymmetry now requires $\sigma_0=0$. Thus,  the supersymmetric 
$adS_d$ solution allowed for
this superpotential is less general than that allowed by a constant 
superpotential. However, we now have the possibility of a 
supersymmetric solution  with non-constant $\sigma$; 
specifically, supersymmetry requires
\be
e^{\alpha\sigma} = \cases{\mp \coth 
\left(\alpha \sqrt{|\Lambda|/2}\, z\right) & $\sigma>0$\cr
\mp \tanh\left(\alpha \sqrt{|\Lambda|/2}\, z\right) & $\sigma<0$\, .} 
\ee
This implies
\be
\alpha\sigma = \pm \log \left|\coth 
\left(\alpha\sqrt{|\Lambda|/2}\, z\right)\right|\, , 
\ee
which is precisely the function $\sigma(z)$ for the $k=0$ 
domain wall solution (\ref{lzero2}), as shown originally in 
\cite{Skenderis:1999mm}.

\item $2\alpha W= \sqrt{2|\Lambda|}\, \sinh \alpha\sigma$. 

This applies for $\Lambda>0$. It is again clear from 
(\ref{susycon}) that only domain wall solutions with
\be
\dot\sigma =  \pm \sqrt{2\Lambda}\, \cosh \alpha\sigma
\ee
can preserve some fraction of supersymmetry. This implies that
\be
e^{\alpha\sigma} = \left|\cot \left(\alpha \sqrt{\Lambda/2}\, 
z\right)\right|\, , 
\ee
which is precisely the function $\sigma(z)$ for the $k=0$ 
domain wall solution (\ref{lzero1}).

\end{itemize}
We have now shown, for $\lambda=0$, that for each flat 
domain wall solution with non-constant 
$\sigma$ there is superpotential for which the supersymmetry 
preserving condition (\ref{susycon2}) is satisfied. 
This implies that the Killing spinor integrability condition 
is satisfied too, and Killing spinors satisfying (\ref{halfsusy}) 
are found by integration of (\ref{intKill}), for $\lambda=0$. 
For the $k=0$ fixed point solution, which is just $adS_d$ 
in horospherical coordinates, there are also Killing eigenspinors
of $\Gamma_z$ with the opposite eigenvalue \cite{Lu:1996rh}, and all
supersymmetries are preserved in this special case. Thus,
{\it all} $k=0$ solutions preserve at least 1/2 supersymmetry  
for some choice of $W$.

\subsection{$\lambda>0$}

When $(\lambda-2\alpha)\Lambda>0$, one possible choice of superpotential is
\be\label{KW}
W= K(\lambda)\, e^{-{1\over2}\lambda\sigma}\, , 
\ee
where $K(\lambda)$ is the function given in (\ref{klambda}). 
For this superpotential the supersymmetry preserving condition 
(\ref{susycon2}) becomes $\dot \sigma = \mp \lambda K(\lambda)$, 
which is satisfied only at the $k=0$ fixed point, as one would 
expect from the fact that an exponential superpotential  is the 
natural generalization to $\lambda>0$ of the constant superpotential 
considered above for $\lambda=0$. One may verify that, for the above
superpotential, (\ref{phisusy}) 
is also satisfied by the $k=0$ fixed point solutions of subsection 
\ref{sec:fixed}, so 1/2 supersymmetry is preserved.  
This is a well-known result in the context of various specific 
supergravity theories with an exponential superpotential. An example 
with $\lambda>\lambda_c$ is the maximal gauged d=8 supergravity 
\cite{Salam:1984ft} for which the $k=0$ fixed point solution was 
shown in \cite{Boonstra:1998mp} to preserve 1/2 supersymmetry. 
There are several cases with $\lambda=\lambda_c$, for $d=5,6,7$. 
An example is the minimal $d=7$ gauged supergravity  
\cite{Townsend:1983kk} for which the $k=0$ fixed point solution was 
shown in \cite{Lu:1995hm} to preserve 1/2 supersymmetry. Cases 
with $\lambda<\lambda_c$ arise from toroidal compactification 
of a higher dimensional model with an $adS$ vacuum \cite{Lu:1996rh}.  
An example for which $\lambda$ is {\it arbitrary} is $d=3$ $adS$ 
N=1 supergravity coupled to a scalar multiplet, the Lagrangian 
and supersymmetry transformation rules  of which can be found  
in \cite{deWit:2004yr}.  

Given our results for $\lambda=0$, it would be natural to suppose 
that there exist $\lambda$-deformations of the $\cosh \alpha\sigma$ 
and $\sinh\alpha\sigma$ superpotentials for which the 
non-fixed-point $k=0$ solutions would also be supersymmetric. To 
investigate this, we differentiate
both sides of (\ref{superpoteq})  with respect to $\sigma$ and 
then use (\ref{superpoteq}) to 
eliminate $\Lambda$. We thus find that the function
\be
X(\sigma) = W'(\sigma)/W(\sigma)
\ee
obeys the first-order ODE
\be\label{ODE1}
2XX' = \left(2X+\lambda\right) \left(\alpha^2-X^2\right)\, . 
\ee
This equation is obviously solved by $X=\pm \alpha$, but this we 
discard because it requires $\Lambda=0$. It is also obviously 
solved by $X=-\lambda/2$; this yields the superpotential 
(\ref{KW}). Unless $\lambda=2\alpha$, all other solutions are given
by\footnote{This is for a specific choice of the integration
  constant, which we may choose without loss of generality;
a change in this constant is equivalent to a shift of
  $\sigma$, which is equivalent to a scaling of
  $\Lambda$, but $\Lambda$ cancels from the ratio $W'/W=X$.
The solution for $\lambda=2\pi$ can be found too, but it is less
illuminating and we omit it.}
\be\label{genX}
\left|X+\lambda/2\right|^{2\lambda} \left| X-\alpha\right|^{2\alpha-\lambda} 
\left| X+\alpha\right|^{-2\alpha-\lambda} = 
e^{(\lambda^2-4\alpha^2)\sigma}\, . 
\ee
For any of the functions $X(\sigma)$ defined implicitly by this
algebraic relation we have the superpotential 
\be
W(\sigma) = \exp\left(\int^\sigma \! X(s)\,  ds\right)\, . 
\ee

For $\lambda=0$, (\ref{genX}) simplifies to 
\be
\left|{X-\alpha\over X+\alpha}\right| = e^{-2\alpha\sigma}\, , 
\ee
and hence 
\be
X= \cases{ \alpha\tanh(\alpha\sigma) & $|X|<\alpha$\cr
\alpha\coth(\alpha\sigma) & $|X|>\alpha$} \qquad \qquad (\lambda=0).
\ee
These yield, respectively, the $\cosh(\alpha\sigma)$ and 
$\sinh(\alpha\sigma)$ superpotentials discussed in the previous 
subsection. Note that in both cases
$|X| \sim \alpha$ as $|\sigma|\to\infty$, and that $X$ has either 
a zero or a pole at $\sigma=0$. 

The implications of (\ref{genX}) when $\lambda\ne0$ depend on the 
sign of  $(2\alpha-\lambda)$:
\begin{itemize}

\item $\lambda<2\alpha$. In this case $\sigma\to\infty$ implies 
either $X\to\alpha$ or $X\to -\lambda/2$, and $\sigma\to -\infty$ 
implies $X\to -\alpha$. There is one solution with $|X|>\alpha$ 
that yields a superpotential with the same asymptotic behaviour 
as the $\lambda=0$ superpotential $W\propto \sinh(\alpha\sigma)$, 
and this superpotential is therefore applicable for $\Lambda>0$.

There are also {\it two} solutions with $|X|<\alpha$, one with 
$X<-\lambda/2$ and the other with $X<-\lambda/2$. These yield 
two superpotentials with the same behaviour for large $|\sigma|$ as the 
$W\propto \cosh(\alpha\sigma)$ superpotential that is applicable 
for $\Lambda<0$.

\item $\lambda>2\alpha$. In this case $\sigma\to\infty$ implies 
$|X|\to \alpha$ and $\sigma\to -\infty$ implies $X\to -\lambda/2$. 
There is one solution with $|X|<\alpha$ and two solutions with $|X|>\alpha$, 
one with $X>-\lambda/2$ and the other with $X<-\lambda/2$.
 
\end{itemize}
Observe that the number of  possible superpotentials is the same as 
the number of possible 
$k=0$ domain-wall solutions, if we ignore the freedom in the choice 
of integration constants. This is no accident, 
as we now demonstrate. 

Given {\it any} of the $k=0$ domain wall solutions of subsection 
\ref{sec:flat}, we have functions 
$\varphi(z)$ and $\sigma(z)$. As long as $\dot\sigma\ne0$, we 
may {\it define} a function of $\sigma$ by
\be
W(\sigma) = F\left(z(\sigma)\right)\, , 
\ee
where 
\be
F(z) = \mp {1\over 2\alpha} e^{-{1\over2}\lambda\sigma(z)}\, \dot\varphi(z)
\ee
and $z(\sigma)$ is the inverse function to $\sigma(z)$. By 
construction, the Killing spinor condition (\ref{phisusy}) is 
satisfied. We now show that the supersymmetry-preserving condition 
(\ref{susycon2})  is also satisfied. Using (\ref{ddphi}) for $k=0$, we see that
\be
\dot F = \pm {1\over2} e^{-{1\over2}\lambda\sigma} \dot\sigma^2\, , 
\ee
and hence that
\be
W' \equiv \dot F/\dot\sigma = \pm {1\over2} 
e^{-{1\over2}\lambda\sigma} \dot\sigma\, . 
\ee
But this is just (\ref{susycon2}). Thus,  every flat domain wall 
for which $\dot\sigma$ is not identically zero determines a putative 
superpotential with respect to which it preserves at least 
1/2 supersymmetry. We say `putative' because we have still to see 
whether the function $W(\sigma)$ that we have defined 
satisfies (\ref{superpoteq}). In fact, this is automatic, as we now
show. The conditions (\ref{susycon2}) and 
(\ref{phisusy}) imply 
\be
W'/W \equiv X= -\alpha u/v\, , 
\ee
and hence 
\be
X' \equiv u^{-1}\dot X = {\dot v \over v} -{\dot u \over uv}\, . 
\ee
But if $(\dot u,\dot v)$ are given by (\ref{ueq}) then this equation 
is equivalent to (\ref{ODE1}), which is itself equivalent to 
(\ref{superpoteq}) for non-zero $W$. Thus, {\it every flat 
domain wall determines a superpotential with respect to which 
it preserves at least 1/2 supersymmetry}\footnote{This conclusion was 
previously arrived at in \cite{Freedman:2003ax}, where it was
also suggested that the construction should apply for {\it arbitrary} 
dilaton potential $V$.}

 \subsection{$k\ne0$}

To complete our analysis, we now consider whether solutions of our
model for non-flat domain walls can also preserve supersymmetry. 
We start from the observation that (\ref{phisusy2}) is equivalent to
\be\label{phisusy3}
W= {\eta\over2\alpha} \sqrt{e^{-\lambda\sigma}\dot\varphi^2
  -{k\over\beta^2}e^{-2\beta\varphi} }
\ee
for some sign $\eta$. Given {\it any} domain wall solution with non-zero 
$\dot\sigma$, we can use this equation to define a putative 
superpotential $W$, using the implicit function $z(\sigma)$
to express $W\left(\sigma(z)\right)$ as a function of $\sigma$. 
As for $k=0$, we can now compute $W'$. Using (\ref{ddphi}) 
to eliminate the $\ddot\varphi$ term, we find that
\be
W' = -{\eta \dot\sigma\dot\varphi
  e^{-{1\over2}\lambda\sigma} \over 2 \sqrt{\dot\varphi^2 - {k\over
      \beta^2}e^{\lambda\sigma -2\beta\varphi}}}\, . 
\ee
This is consistent with (\ref{susycon2}) iff
\be
\dot\sigma\left[\dot\varphi \pm \eta \sqrt{\dot\varphi^2 - {k\over
      \beta^2}e^{\lambda\sigma -2\beta\varphi}}\right]=0\, .
\ee
As we have assumed that $\dot\sigma$ is non-zero, this implies that
$\eta= \mp 1$ and that $k=0$, in which case (\ref{phisusy3}) reduces
to (\ref{phisusy}). 

Thus, no non-flat dilaton domain walls with non-constant $\sigma$ can be
``supersymmetric'' in the sense of this paper. Supersymmetric
$k=-1$ dilaton domain walls have been found in $d=5$ supergravity
theories \cite{Cardoso:2002ec,Behrndt:2002ee}, but these involve
additional scalar fields. The possibility of supersymmetric $k=-1$
domain walls in models with a single scalar field has been studied 
in \cite{Freedman:2003ax}, via the introduction of an $su(2)$-valued
superpotential, with the conclusion that the $adS_{d-1}\times
\bR$ solution (obtained here as the $k=-1$ fixed point solution 
for $\Lambda<0$ and $\lambda=0$) is ``fake supersymmetric''. 
The phase-plane analysis shows that there exists 
a solution that interpolates between the $adS_d$ and $adS_{d-1}\times\bR$
fixed-point solutions.
We found  this ``separatrix wall'' solution exactly for $d=3$. There are
actually four separatrix wall solutions, corresponding to the four 
possible trajectories that connect an $adS_d$ fixed point to an 
$adS_{d-1}\times \bR$ fixed point. This is illustrated for $d=3$ in 
Fig. 6a.

\begin{figure}[!h]\label{diamond}
\vskip2em
  \begin{center}
    (a)\epsfig{file=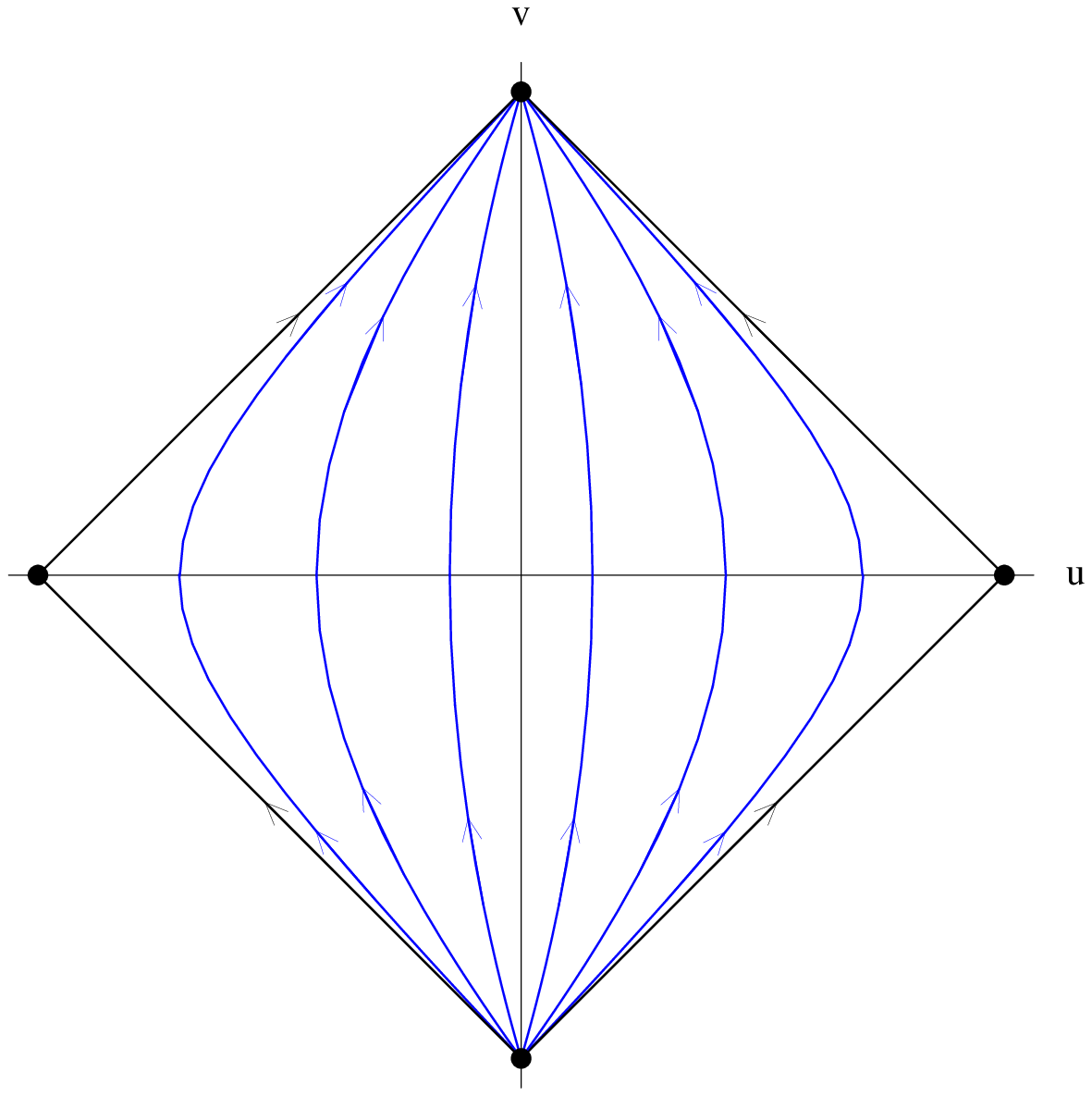,width=6.5cm}\hskip1em (b)\epsfig{file=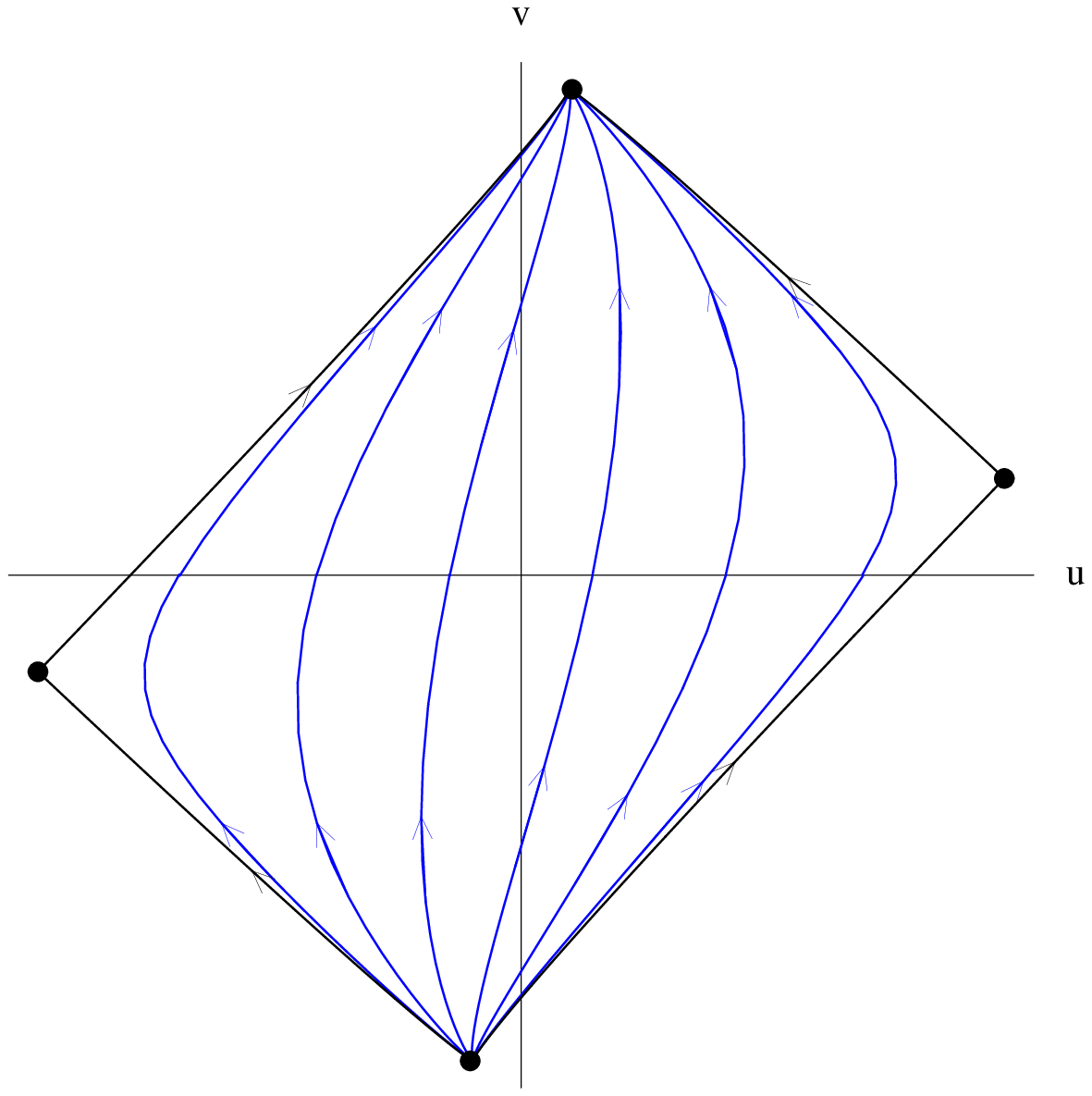,width=6.5cm}\hskip3em\begin{picture}(2,2)(0,0)
      \put(-145,205){$\Lambda<0$,\, $\lambda =0$}
      \put(60,205){$\Lambda<0$,\, $0<\lambda < \lambda_c$}
     \end{picture}
     \end{center}\caption{\small{(a) The central region of Fig. 2a for
      $d=3$. The ``Janus'' trajectories interpolating
      between the two $adS_3$ fixed points may come arbitrarily close to
      one of the $adS_2\times\bR$ fixed points. In the limit one gets
      a trajectory that is the union of two trajectories, each corresponding
      to the exact ``separatrix wall'' solution of section
      \ref{sec:exact}.}. (b) The small $\lambda$ deformation of the
      Janus and separatrix trajectories.}
\end{figure}

The phase-plane analysis also shows that there exists a one-parameter 
family of solutions that interpolates between two isometric 
$adS_d$ spacetimes. These are the Janus solutions discussed in 
\cite{Freedman:2003ax}. There is a Janus solution with a phase-plane 
trajectory that approaches arbitrarily close to the union of two
separatrix trajectories. This can be viewed as a marginal bound state
of two separatrix walls, separated by an arbitrary distance (related to the 
parameter $c$ in section \ref{subsection}). This is a rather 
unusual state of affairs, reminiscent of the multiple domain wall 
solutions  of certain supersymmetric sigma-models
\cite{Gauntlett:2000ib}. It suggests a
no-force condition that is usually associated with supersymmetry.
Indeed, it is claimed in \cite{Freedman:2003ax} that the Janus solutions 
{\it are} (``fake'') supersymmetric, and continuity would then 
suggest that the separatrix walls have the same property. 

As shown in Fig 6b, the same considerations
apply for any $\lambda<\lambda_c$ in that there still exists a
one-parameter family of Janus-type solutions that are asymptotic to
both $k=0$ fixed point solutions, but with the difference that these
fixed point solutions are no longer $adS$ spaces.

\section{Comments}

We have shown in this paper how the equations governing domain 
wall solutions of $d$-dimensional gravity coupled to a dilaton 
with an exponential potential define a family of 
2-dimensional autonomous dynamical systems, parametrized by the dilaton 
coupling $\lambda$, and with a transcritical bifurcation as a 
function of $\lambda$ when the dilaton potential is negative. 
This formulation of the problem, which is analogous to the similar 
formulation of homogeneous and isotropic cosmologies in the same class of models, 
allows a much more complete understanding of the space of domain-wall 
solutions than has hitherto been possible, particularly for curved 
domain walls for which the worldvolume geometry is de Sitter ($k=1$) 
or anti de Sitter ($k=-1$).

One difference with the cosmological case is that domain walls 
can preserve some fraction of supersymmetry, and we have shown that
{\it all} flat walls are ``supersymmetric'' with respect to some 
superpotential $W$ for which $d\log W$ can be found exactly, 
albeit implicitly, for any $\lambda$. Of course, whether this
superpotential actually arises in the context of some supergravity 
theory is another question, and one that we have not addressed in any
detail. We note however that any superpotential is possible in $d=3$
and that there therefore exist supersymmetric domain walls in $d=3$
supergravity models with $\lambda=2\alpha$, or $\Delta=0$ in the 
notation of \cite{Lu:1995hm,Lu:1996rh}; this is a case for 
which no domain wall solution, supersymmetric or otherwise, 
was previously known.

In the case of a pure cosmological constant, corresponding to 
$\lambda=0$, we found the exact phase-plane trajectories. For 
$\Lambda<0$ there are two $k=-1$ fixed points, each corresponding to 
an $adS_{d-1}\times \bR$ solution found in \cite{Freedman:2003ax};  
this is the analog of the Einstein Static Universe that 
occurs for $\Lambda>0$ in the cosmological case. There are also 
(isometric) ``separatrix wall'' solutions  with phase-plane trajectories
that interpolate between an $adS_{d-1}\times \bR$ fixed 
points and one of two $adS_d$ $k=0$ fixed-points; we found the 
exact separatrix trajectories, and the exact solution for $d=3$. 

The trajectories that interpolate between the two $adS_d$ fixed points 
correspond to the one-parameter family of ``Janus'' solutions of
\cite{Freedman:2003ax}, which we have interpreted as marginal bound
states of two Separatrix Walls (and the same applies to the ``deformed'' Janus 
solutions for $0<\lambda<\lambda_c$). This interpretation makes physical 
sense for large separation, but for zero separation the solution
degenerates to the $adS_d$ vacuum.  This suggests a more precise 
interpretion of the Janus solutions as separatrix wall/anti-wall bound 
states. It might seem unlikely that such a configuration could be stable
(as is shown in \cite{Freedman:2003ax}) but the usual intuition need not apply 
in an $adS$ (or deformed $adS$) background.

\vskip 1cm
\noindent
{\bf Acknowledgements}:
We are grateful to Eric Bergshoeff,  Andres Collinucci and Diederik 
Roest for the discussions on domain walls that led to the work 
described here, and for allowing us to take over unchanged some notation 
and preliminary results from their unpublished notes on this topic.
We also thank Mirjam Cveti{\v c} and Kostas Skenderis for helpful
correspondence. J.S. thanks the following bodies for financial
support: the Gates Cambridge Trust, der Studienstiftung des deutschen 
Volkes, Trinity College Cambridge and PPARC.

\end{document}